\documentclass[11pt,a4paper]{article}
\pdfoutput=1
\usepackage{jheppub}

\usepackage{xspace}
\usepackage{epsfig}
\usepackage{amssymb,amsmath}
\usepackage{url,booktabs}
\usepackage{color}
\usepackage{slashed}
\usepackage{subfig}
\usepackage{xfrac}
\usepackage{wrapfig}
\usepackage{tikz}
\usepackage{booktabs}
\def\be{\begin{equation}}
\def\ee{\end{equation}}
\def\bea{\begin{eqnarray}}
\def\eea{\end{eqnarray}}
\def\beal{\begin{align}}
\def\eeal{\end{align}}
\def\checkmark{\tikz\fill[scale=0.4](0,.35) -- (.25,0) -- (1,.7) -- (.25,.15) -- cycle;}

\newcommand{\Dplus}{\Delta_{+}}
\newcommand{\mchi}{m_{\chi}}
\newcommand{\met}{ $\slashed{E}_T~$}
\newcommand{\metm}{\slashed{E}_T}

\newcommand{\ra}[1]{\renewcommand{\arraystretch}{#1}}

\graphicspath{{Plots/}}
\relax
\title{Closing the window for compressed Dark Sectors with disappearing charged tracks.}

\author[a,b]{Rakhi Mahbubani}
\author[c]{Pedro Schwaller}
\author[d,e]{and Jos\'e Zurita}

\affiliation[a]{Theoretical Physics Department, CERN, Geneva, Switzerland}
\affiliation[b]{Theoretical Particle Physics Laboratory, Institute of Physics, EPFL, Lausanne, Switzerland}
\affiliation[c]{PRISMA Cluster of Excellence \& Mainz Institute for Theoretical Physics, Johannes Gutenberg University, 55099 Mainz, Germany}
\affiliation[d]{Institute for Nuclear Physics (IKP), Karlsruhe Institute of Technology, Hermann-von-Helmholtz-Platz 1, D-76344 Eggenstein-Leopoldshafen, Germany}
\affiliation[e]{Institute for Theoretical Particle Physics (TTP), Karlsruhe Institute of Technology, Engesserstra{\ss}e 7, D-76128 Karlsruhe, Germany} 
\emailAdd{rakhi@cern.ch}
\emailAdd{pedro.schwaller@uni-mainz.de}
\emailAdd{jose.zurita@kit.edu}

\abstract{We investigate the sensitivity at current and future hadron colliders to a heavy electrically-charged particle with a proper decay length below a centimetre, whose decay products are invisible due to below-threshold energies and/or small couplings to the Standard Model.
A cosmologically-motivated example of a framework that contains such a particle is the Minimal Supersymmetric Standard Model in the limit of pure Higgsinos.
The current hadron-collider search strategy has no sensitivity to the upper range of pure Higgsino masses that are consistent with the thermal relic density, even at a future collider with 100~TeV centre-of-mass energy.
We show that performing a disappearing track search within the inner 10 cm of detector volume would improve the reach in lifetime by a factor of 3 at the 14~TeV LHC and a further factor of 5 at a 100~TeV collider, resulting in around 10 events for 1.1~TeV thermal Higgsinos. In order to include the particles with the largest boost in the analysis, we furthermore propose a purely track-based search in both the central and forward regions, each of which would increase the number of events by another factor of 5, improving our reach at small lifetimes.
This would allow us to definitively discover or exclude the experimentally-elusive pure-Higgsino thermal relic at a 100~TeV collider.}

\arxivnumber{1703.abcde}
\notoc

\begin{document}
\begin{flushright}
CERN-TH/2017-054  \\ TTP17-005 \\ MITP/17-011
\end{flushright}
\maketitle
%-----------------------------------------------------------------------------
\section{Introduction}
\label{sec:intro}
%-----------------------------------------------------------------------------

In spite of the continued shattering of performance goals at the LHC, these are unexpectedly lean times for particle phenomenologists.  The wealth of interesting excesses over the Standard Model (SM) that were
anticipated, and indeed prepared for, in the run-up to data-taking,
have failed to materialise.  In their absence the attention of some
part of the community has turned to possible signatures of
new physics that may be overlooked by our current experimental
program. On closer examination this avenue has been extremely
fertile, and is now seen to be linked to
some of the more exotic
examples of Physics Beyond the Standard Model, such as Hidden
Valleys~\cite{Strassler:2006im}, or Neutral Naturalness~\cite{Chacko:2005pe,Burdman:2006tz}, ideas that are
currently enjoying a renaissance as a result of our reduced
circumstances.

One does not need to resort to exotic models, however, to find
signatures that fall through holes in our experimental net.
Even our favourite scenarios for physics Beyond the Standard Model
(BSM) can yield these, in regions of parameter space that remain relatively
unexplored due in no small part to a strong theory prior.  As our
biases are eroded by data, this less-charted territory merits a more thorough exploration to ensure these gaps in coverage are filled in. A 100-TeV collider is, in that sense, \emph{tabula
rasa}; it provides us the opportunity to build in sensitivity
for unconventional signals that are an afterthought at
established experimental programs.

An example receiving some attention of late is compressed scenarios, with lowest-lying new states that are
near-degenerate in mass.  This could lead to long-lived intermediates through
kinematic suppression of decay widths, as well ad decay by-products that are below the energy threshold for detection.  One means of probing such states is by direct measurement of the track due to an electrically-charged parent, with accessible particle lifetimes set by the characteristic size of the detecting apparatus.  Hence, unlike the Energy and
Intensity frontiers, pushing this `Lifetime
Frontier'~\cite{Chou:2016lxi} does not necessarily require the design and construction of an
entire experimental facility, but rather moves the focus to detectors
and their design; either through the building of a single-purpose long-lived particle detector, as proposed
in~\cite{Chou:2016lxi}, or adapting a multipurpose detector so it is better
suited to measuring a larger range of lifetimes.  It is the latter
that concerns us here, specifically in relation to
electrically-charged particles that are too short-lived ($c\tau\lesssim
\mathcal{O}$(cm)) to leave a conventionally-measurable track at current particle
detectors.  

The investigation of intermediate-lifetime particles is by no means a modern
endeavour.  Indeed much of the BSM physics program at the LHC relies on being able
to tag a bottom quark due to the mm-scale decay length of the
$B$-hadron it results in. In this
case, however, the existence of the $B$-hadron is inferred by
extrapolation of the tracks of its decay products to a displaced vertex, rather than a
measurement of the properties of the parent particle itself.  The
latter will be indispensable to probe
examples where the daughter particles are
undetectable, either because they have very little energy, or because
they are non-interacting components of a `dark
sector'.  

A canonical example of a model containing such a particle
is the decidedly un-exotic Minimal Supersymmetric Standard Model (MSSM),
whose neutralino sector provides various well-motivated candidates for
thermal relic dark matter.  Pure Wino and Higgsino states
have charged components that are heavier than neutral ones by
electroweak loop effects; these $\mathcal{O}$(100~MeV) splittings
result in $c\tau\sim$ mm-cm.  Recent
work~\cite{Low:2014cba,Cirelli:2014dsa} has shown the entire range of viable thermal Wino masses to be discoverable at a 100~TeV collider in this channel.  However this is not the case for pure
Higgsinos, which have charged-neutral splittings that are twice as
large, the decay width being strongly dependent on the splitting.
Pure Higgsino dark matter is also particularly difficult to access
directly by other means, since its tiny indirect and direct detection cross
sections are beyond even the projected sensitivity of any dark matter
experiment currently under consideration. 

In this work, we explore the dependence of the reach for such
intermediate-lifetime charged particles, on the tracker properties at a
hadron colliders, using the disappearing track signature.\footnote{For recent work on long-lived electrically charged particles at the LHC, see~\cite{Ostdiek:2015aga,Khoze:2017ixx}.} Unlike
many existing searches for compressed electroweak-charged states \cite{Giudice:2010wb,Gori:2013ala,Han:2013usa,Schwaller:2013baa,Baer:2014cua,Han:2014kaa,Bramante:2014dza,Han:2014xoa,Baer:2014kya,Gori:2014oua,Liu:2014lda,Bramante:2014tba,Han:2015lma,Barducci:2015ffa,Badziak:2015qca},
we operate under the assumption that no information can be obtained
from their decay products, making us less sensitive to
the origin and properties of the parent. We then
express our results in the parameter space of thermal Higgsino dark
matter, and show that full coverage of the elusive pure Higgsino
region ($m_{\chi} \lesssim 1.1~\rm{TeV}$) can be achieved with a total
integrated luminosity of 3000 fb$^{-1}$.  While our main 
focus is a 100~TeV proton-proton collider (FCC-hh), we also examine similar
upgrades to ATLAS and CMS that could improve the
 LHC reach for compressed Higgsinos at its
high-luminosity run (LHC14-HL).   In a
companion paper \cite{Mahbubani:2017tba} we study the reach in the di-lepton plus missing
transverse energy channel,
which doesn't assume the presence of an electrically-charged state,
but relies instead on additional weak radiation from the
initial state, in the form of a leptonic $Z$-boson. 

\section{Simplified model}
\label{sec:SimpMod}
\begin{figure}[t]
\begin{minipage}[c]{0.47\textwidth}
\center
    \begin{tikzpicture}[baseline=(current bounding box.north),line width=1 pt, scale=0.75]
      \draw (0,0) -- (-2,1.5) node [inner sep = 2mm, left] {p};
      \draw (0,0) -- (-2,-1.5) node [inner sep = 2mm, left] {p};
      \draw (0,0) -- (3,1.5) node [inner sep = 2mm, above right] {Invisible};
      \draw (0,0) -- (3,-1.5) node [inner sep = 2mm, below right] {Invisible};
      \node at (0,0) [circle, fill=gray,thick, minimum
      size=1.5cm]{};
      \node [inner sep = 0.5mm,circle,fill=black] (vertexup) at (2,1) {};
      \node [inner sep = 0.5mm,circle,fill=black] (vertexdown) at (2,-1) {};
      \draw[dashed] (vertexup) -- (50:3) ;
      \draw[dashed] (vertexup) -- (45:3) ;
      \draw[dashed] (vertexup) -- (40:3) ;
      \draw[dashed] (vertexdown) -- (-40:3) ;
      \draw[dashed] (vertexdown) -- (-45:3) ;
      \draw[dashed] (vertexdown) -- (-50:3) ;
      \draw[white] (-1.2,1) -- (0.5,2) node [inner sep = 2mm, above ] {$j$/$Z$/$\gamma\cdots$};
      \node [inner sep = 2 mm,above] at (1.2, 0.5) {$\chi^+$};
      \node [inner sep = 2 mm,below] at (1.2,-0.5) {$\chi^-$};
      \draw[magenta] (0.85,0.43) -- (vertexup);
      \draw[magenta] (0.85,-0.43) -- (vertexdown);
    \end{tikzpicture}
    \end{minipage}\hfill
    \begin{minipage}[c]{0.47\textwidth}
    \vspace*{2\baselineskip}
\caption{Production of a charged state with proper lifetime
  $\tau\lesssim 1$ ns and decay products that are invisible at colliders will lead to a charged track that ends (`disappears') within the extent of a tracker subsystem.}
\label{fig:SimpChargedTrack}
\end{minipage}
\end{figure}
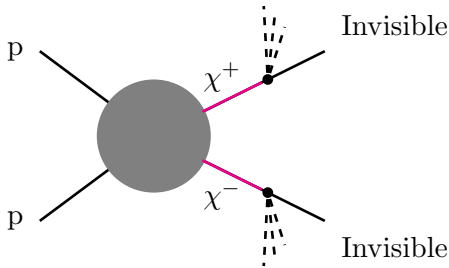
Our disappearing track search will be relevant to any scenario
containing a charged particle with proper lifetime $\tau$ below 10
picoseconds, and whose decay products are invisible, either due to
small energies or small couplings to the SM, see
Fig~\ref{fig:SimpChargedTrack}.  Such states are too
short-lived to be covered by conventional disappearing track searches at
current \cite{Aad:2013yna,CMS:2014gxa} or future \cite{Low:2014cba} colliders.  
We attribute to the charged state a `nominal decay length' $c\tau$, which translates into an average lab-frame decay length of $\beta\gamma c\tau$ for a particle with velocity $\beta=v/c$ and Lorentz boost $\gamma$.  Converting this to an actual charged track length requires us to take into account the Poissonian nature of the decay process, and weight the decay length by the probability that the chargino will travel a distance $d$ without decaying, given by
\be
\mathcal{P}(d)=\exp{\left(-\frac{d}{\beta\gamma c\tau}\right)}\;.
\ee
We carry out our simulation and analysis within a specific framework containing such a particle, where the usual Standard Model field content is supplemented with a new vector-like weak-doublet fermion with hypercharge $\sfrac{1}{2}$, $\chi=\left( \chi^+ ,\chi^0\right)$ and Dirac mass $
m_\chi$:   
\begin{equation}
\mathcal{L}\supset i \bar{\chi}\left(\slashed{\partial}-i
  g\slashed{W}-i g^\prime\frac{1}{2}Y\slashed{B}\right)\chi + m_\chi \bar{\chi}\chi \;,
\end{equation}
where we assume that the neutral component $\chi^0$ is stable on collider timescales.  Electroweak loops will raise the mass of the charged component,
giving rise to a charged-neutral splitting
$\Delta_+=m_{\chi^+}-m_{\chi}\sim \mathcal{O}(\alpha m_Z)$.  The precise value of this one-loop effect is specific to the quantum numbers of the multiplet in question, and for the weak doublet is $\sim 360$~MeV for $m_Z \ll m_\chi$ \cite{Thomas:1998wy,Cirelli:2005uq}, yielding a nominal decay length of 6.6 mm.  This corresponds
to the pure Higgsino limit of the MSSM, with all other superparticles
decoupled.
  
The decay width of the charged component of the doublet is strongly
dependent on the splitting $\Delta_+$, as is evident from the
expressions for the partial decay widths in Appendix \ref{sec:Decay}.
Since we focus on the case where any SM by-products of the
decay are too soft to be detected, we are insensitive to the identity
of the daughter particles.  As a result we simply allow the nominal
decay length of the charged component to be a free parameter, in order
to take into account possible additional contributions to the
splitting from couplings heavy states.\footnote{These could be
  parametrised, for example, by the Wilson coefficient of a
  dimension-5 operator $\bar{\chi}\sigma^a\chi H^\dag\sigma^a H$ for
  Pauli matrices $\sigma$, coming from integrating out a heavy
  weak-triplet scalar.}  A proper lifetime of 1-10 picoseconds
corresponds to a nominal decay length in the mm-cm range.

If we further assume that the neutral component $\chi_0$ is stable
on cosmological timescales, $\chi$ would be natural thermal relic,
saturating the measured relic densirt
density for a mass $m_{\chi} \approx 1.1$~TeV.\footnote{We would also
  need to introduce a small Majorana splitting between the two neutral
  components in order to remain consistent with null results from
  direct detection experiments, a vector-like coupling with the $Z$ being
  long excluded.  However the necessary splitting is small compared with the charged-neutral splitting $\Delta_+$, and as such can be neglected when considering the relic density and collider phenomenology.}  In this work we will remain agnostic about the dark matter connection, and scan over masses beyond this range, although we will use the parameters $m_\chi=1.1$~TeV and $c\tau=6.6$ mm as references throughout.

\begin{figure}
\subfloat[]{
\includegraphics[width=0.49\textwidth]{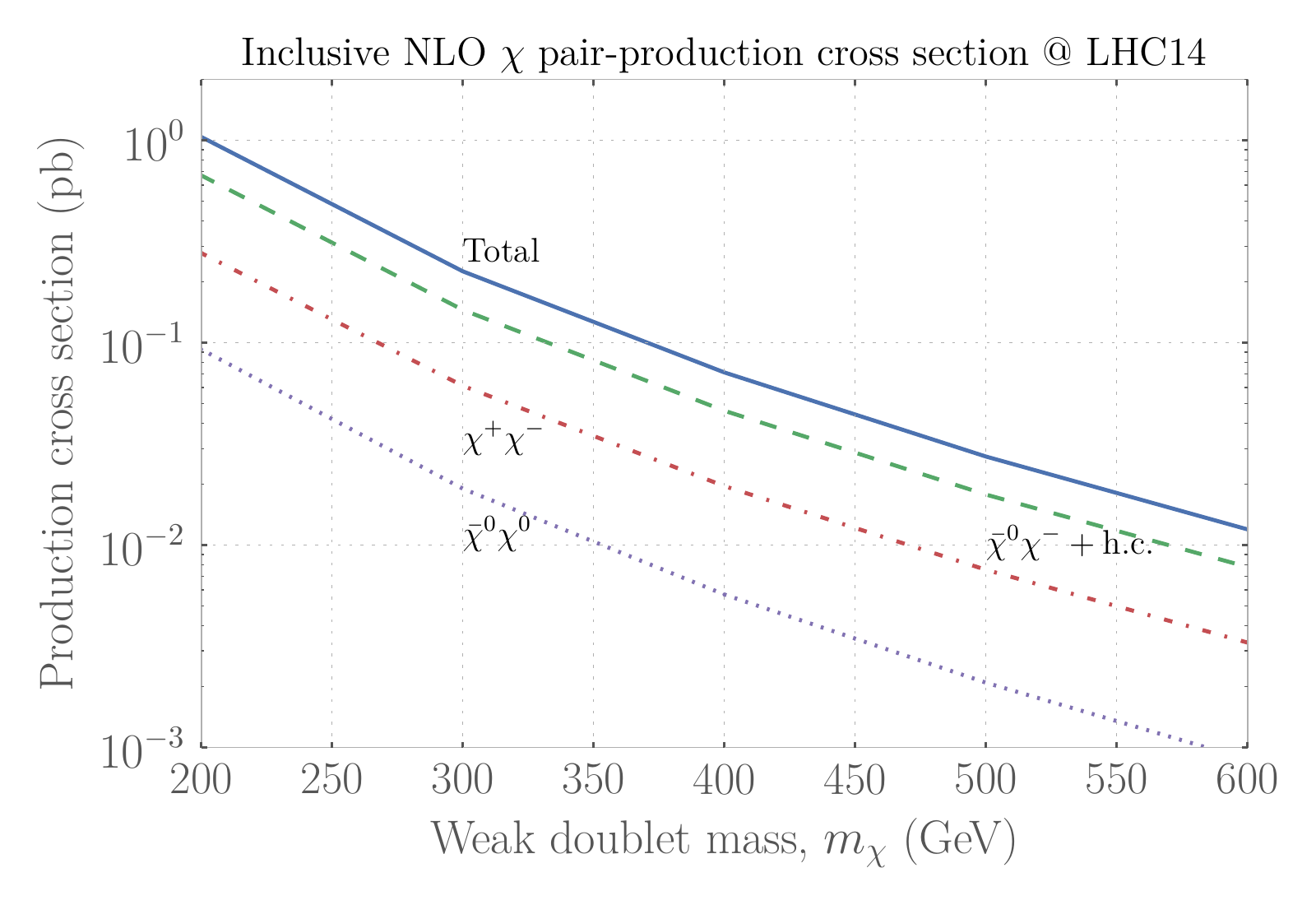}}
~
\subfloat[]{
\includegraphics[width=0.49\textwidth]{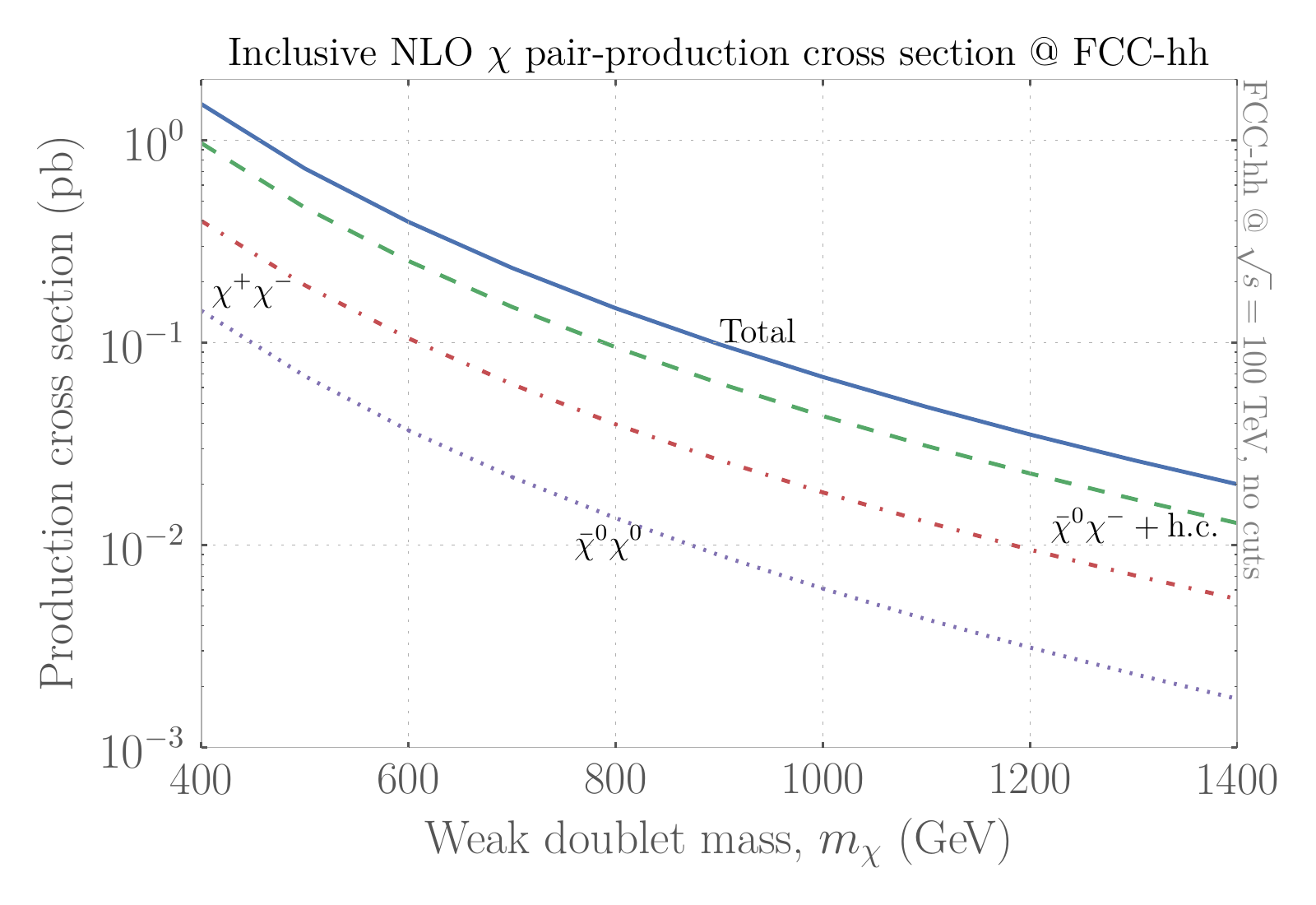}}
\caption{Pair-production cross sections for a weak doublet with hypercharge $\sfrac{1}{2}$ at NLO, obtained using
   \texttt{PROSPINO} in the pure Higgsino limit, for LHC14 (a) and FCC-hh (b).}
  \label{fig:NLOXS}
\end{figure}

We simulate inclusive pair-production of weak doublet $\chi$ using MadGraph5\_aMC@NLO \cite{Alwall:2014hca}, matching up
to two extra jets using the MLM scheme \cite{Mangano:2002ea}, and
shower using Pythia 6.4 \cite{Sjostrand:2006za} with $k_t$-showers
turned on.  Events were  clustered with FastJet \cite{Cacciari:2011ma} and analyzed with the help of MadAnalysis 5
\cite{Conte:2012fm,Conte:2014zja,Dumont:2014tja}.
We normalize the production rate to the NLO cross sections computed
with  \texttt{PROSPINO} \cite{Beenakker:1999xh} for pure Higgsino pair
production; these are shown in Fig~\ref{fig:NLOXS} for the LHC14 (a) and FCC-hh (b).

%-----------------------------------------------------------------------------
\section{Disappearing charged tracks}
\label{sec:events}
%-----------------------------------------------------------------------------

It is clear from the outset that the boost of the charged particle will be a crucial factor in
detecting states with nominal decay length below a
centimetre. Indeed, if this particle is directly produced,
reconstructing its track using tracking elements situated at tens of
centimetres from the interaction point would require on average a boost factor $\gamma\beta\sim\mathcal{O}(100)$ were the Poissonian nature of the
decay process not taken into account.  As it stands, with such a
large centre-of-mass energy at our disposal, neither large boost nor
large statistics should be a limiting factor for TeV-scale particle searches at 100~TeV centre-of-mass.
%%%%%%%%%%%%%%%%%%%%%% Fig 1 %%%%%%%%%%%%%%%%%%%%%%%%%%%%%
\begin{figure}[htb]
\subfloat[]
{%
\includegraphics[width=0.49\textwidth]{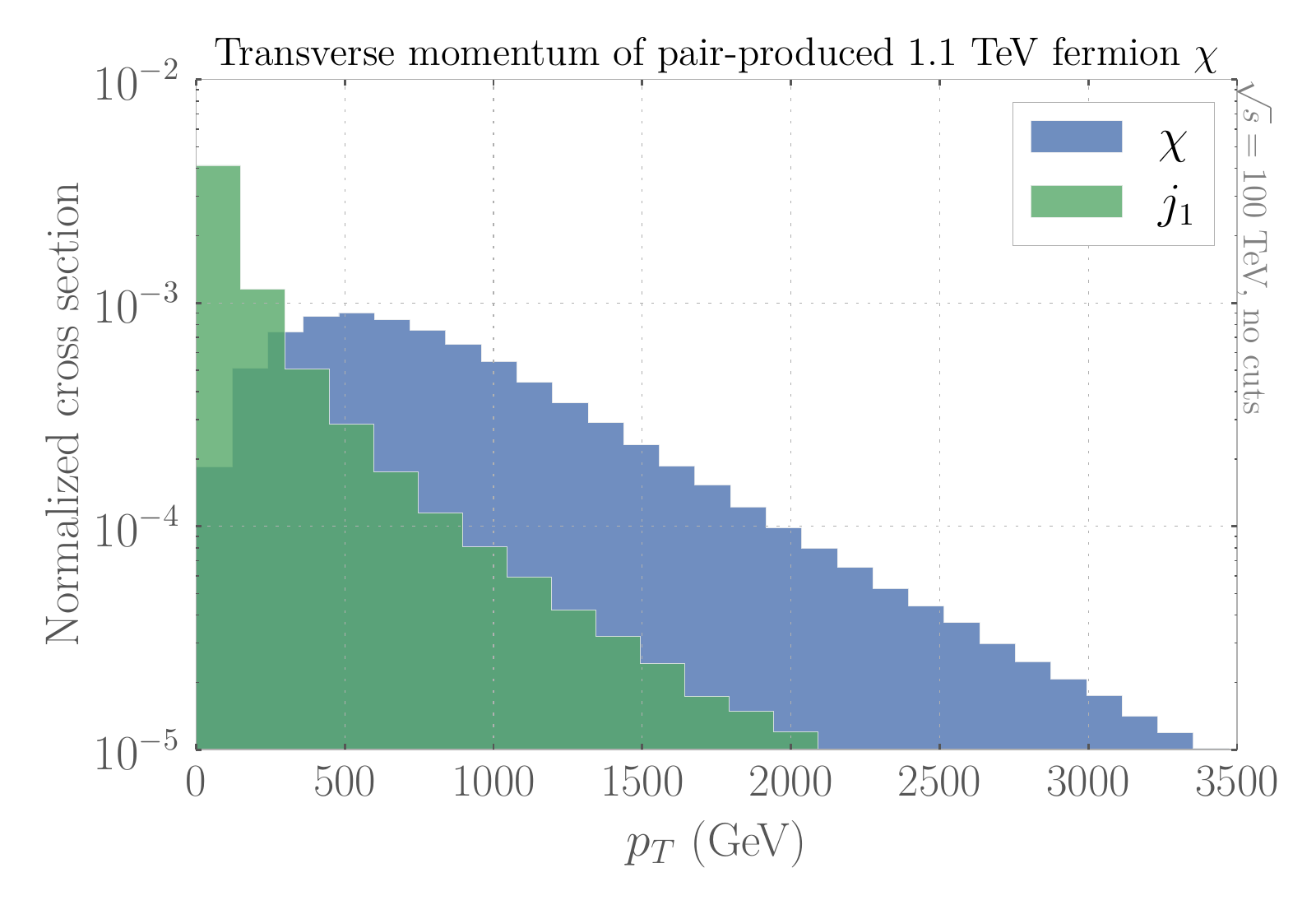}}
~
\subfloat[]
{
\includegraphics[width=0.49\textwidth]{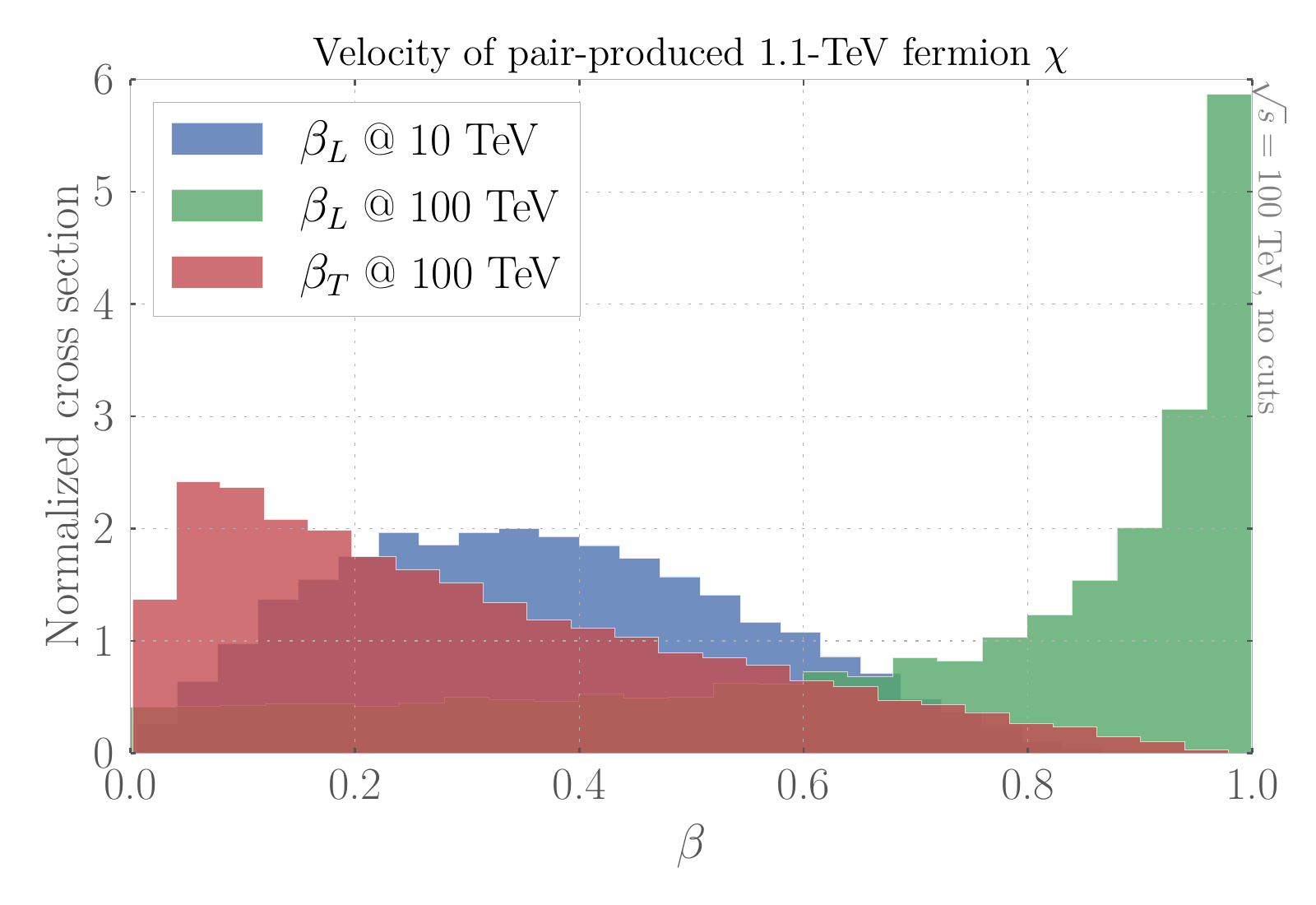}}
\\
\subfloat[]
{%
\includegraphics[width=0.49\textwidth]{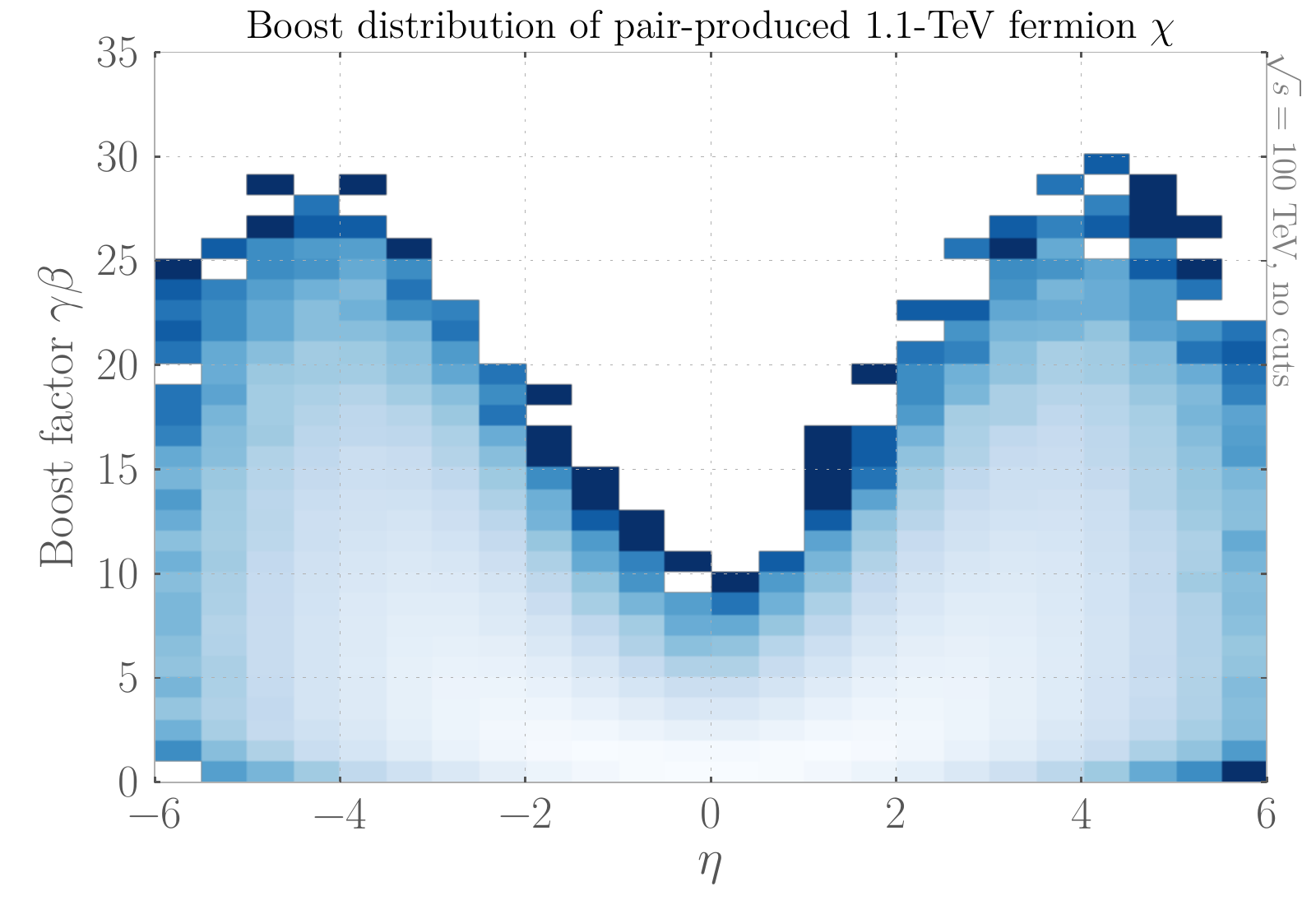}}
~
\subfloat[]
{%
\includegraphics[width=0.49\textwidth]{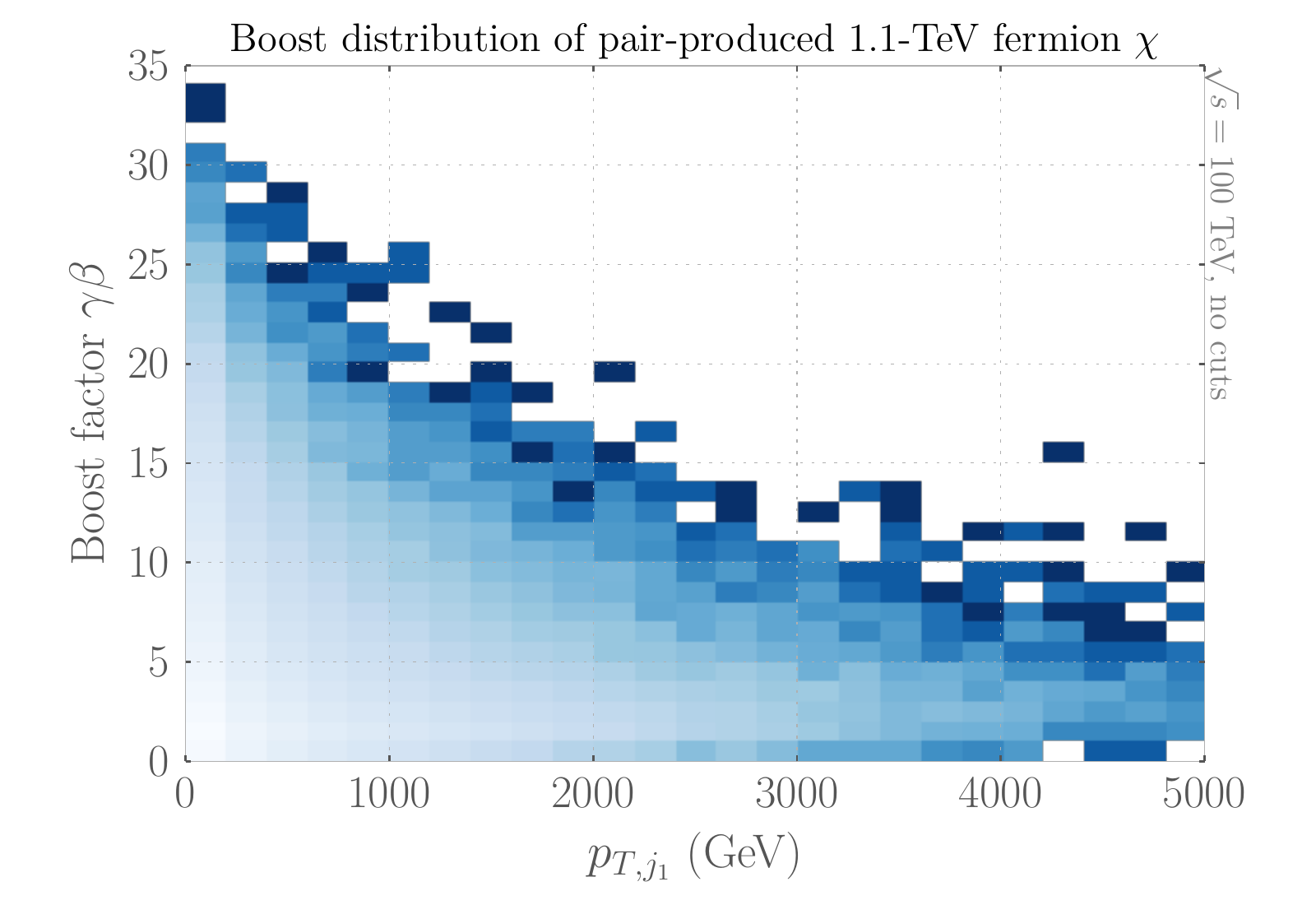}}
\caption{Kinematic distributions (normalised to unity) for
  pair-produced weak-doublet fermions $\chi$ with Dirac mass $m\chi=1.1$~TeV at FCC-hh. We
  show 1-dimensional histograms for (a) the $p_T$ spectra
  of the pair-produced particle and leading jet and (b) the transverse
  and longitudinal beta distributions ($\beta_T$ and $\beta_L$) of the
  pair-produced particle, with
  $\beta_L$ at 10~TeV centre-of-mass for comparison.  We also show
  correlations between the boost factor $\gamma\beta$ of $\chi$ and
  (c) $\chi$ pseudorapidity $\eta$ and (d) the $p_T$ of the
  leading jet, with lighter regions signalling a higher density.  Details of the
  model and simulation can be found in the text.}
\label{fig:MomDistrib}
\end{figure}
%%%%%%%%%%%%%%%%%%%%%%%%%%%%%%%%%%%%%%%%

There is an inherent limitation, however, on the size of the transverse boost:
the transverse momentum of a pair-produced particle at a hadron
collider is roughly set by the particle mass, as illustrated in
figure~\ref{fig:MomDistrib}(a) for $m_\chi=1.1$~TeV.  By contrast, we
can infer from the velocity ($\beta$) distributions in
figure~\ref{fig:MomDistrib}(b) that the particle mass is negligible on
the scale of the longitudinal momentum, resulting in charged particles
with near-relativistic longitudinal velocities ($\beta_L$).  The far more
modest transverse velocity ($\beta_T$) distribution, as well as the
$\beta_L$ distribution for non-negligible mass, are shown for
comparison.  Hence for a fixed nominal decay length, the particles with the
longest lab-frame decay length, corresponding to the those with the
largest boost, are predominantly found in the forward
direction, see figure~\ref{fig:MomDistrib}(c), in a pseudorapidity
region which we will see is not covered in existing disappearing track
searches.  The boosts in the
central region are more moderate, although the particle density is
larger (lighter region in the figure).  Both these
properties will play an important role in our analysis.

The more challenging problem for existing disappearing track searches,
which have generic dark matter searches at their core, becomes
immediately apparent on examining figure~\ref{fig:MomDistrib}(d).  With little else
to work with, in events with pair-produced dark-sector particles we usually
rely on a visible radiated particle $X$ (could be a
jet, photon, $Z$-boson, Higgs,...) recoiling off the dark pair with
large $p_T$. This approach is counterproductive in a disappearing
track search.  In requiring a minimum recoil $p_T$, we
are effectively excluding from the analysis those charged particles 
with the longest lab-frame decay length: for a fixed available energy, the
more energetic the recoil state the less energy is available for boosting
the charged particles.\footnote{Although the above distributions were obtained using the simplified model detailed in Section~\ref{sec:SimpMod} above, we expect their salient features to be mostly
dependent on kinematics, and hence relatively insensitive to the
particulars of the model used.}  To
fully benefit from the availability of highly-boosted
particles will require a radical change
in the philosophy of disappearing track searches, which currently rely
almost exclusively on $X$ $p_T$ (and related quantities such as
missing transverse energy (MET).

In order to discard recoil $p_T$-based quantities a viable alternative must be
found.  Fortunately the most interesting events in this particular
search have a characteristic feature which is rather unusual: a
charged track with extremely high momentum, a quantity which could
potentially be used as replacements for jet $p_T$.  Track $p_T$ is already used to
define signal regions in the current ATLAS search~\cite{Aad:2013yna},
but this measurement could also be combined with track pseudo-rapidity
to obtain total track momentum.  By setting the boost of the
charged particle, track momentum is also directly correlated
with the average track length in the lab frame. 

However it is impossible to know how precisely this quantity
can be measured, particularly for short tracks, or whether this
measurement can be made `on-the-fly'. Moreover, since
there exist electrically-charged QCD bound states with similar
lifetimes, the hadronic background in the absence of a MET cut will
likely be not insignificant.  Since we currently have no data on the
momentum spectra of such backgrounds, we will be unable to
give a
quantitative assessment of the sensitivity of an analysis based on this
quantity.  However we can still make a rough comparison with a more
conventional analysis on the basis of numbers of signal events, keeping in mind
that each analysis selects a subset of events with different properties, and
hence different backgrounds, both with respect to size and composition.

In this work we will pursue two strategies: we begin with a more conventional search
modelled loosely on the jet-$p_T$-based disappearing charged track analysis
carried out by the ATLAS collaboration at 8~TeV~\cite{Aad:2013yna} with rescaled
cuts.\footnote{The CMS analysis \cite{CMS:2014gxa} has a higher mass reach, but
less sensitivity at small lifetimes.}  Our second approach is more
speculative, employing a track-based analysis which will assume the successful
use of some combination of charged-track momentum, and perhaps also $dE/dx$, to
 select potentially interesting events in a kinematic
regime where they are not swamped by Standard Model QCD
backgrounds.\footnote{We thank Phil Harris and Maurizio Pierini for
  bringing to our attention the possibility of using $dE/dx$ as a
  discriminating variable.}  The exact value of the momentum trigger is not crucial
here, since the charged particles with largest momenta have a higher
probability of surviving to large distances, but for the purposes of this analysis we assume
$|p|_{\rm track}\gtrsim$ a few TeV will suffice. In each case we will
explore the effect that varying the detector setup has on the
sensitivity of the analysis in question. The detailed event selection
for each study is summarized in table \ref{tab:cuts}.

\begin{table}
\centering
\ra{1.4}
\begin{tabular}{@{} c c c|| c c c @{}}
\toprule
   & \multicolumn{2}{c}{LHC14-HL}& \multicolumn{3}{c}{FCC-hh }
                                       \\
  \cmidrule{2-3}\cmidrule{4-6}
   & Conventional & \multicolumn{1}{c}{TB Central}     & Conventional   & TB Central & TB Forward\\
\midrule
 lepton veto &  \checkmark  & \checkmark & \checkmark & \checkmark & \checkmark\\
$p_{T,j_1}$ (GeV) &  150    & & 1000 & &\\
\met (GeV)&  150  & & 1000 & &\\
$\Delta\phi^{j,\metm}_\text{min}$ & 1.5 &  & 1.5 & \\
$p_{T,\text{tr}}$ (GeV) & 350  & 1000 & 1000 & 3000 & \\
$p_\text{tr}$ (GeV)& & & & & 8000 \\
$|\eta_\text{tr}|$ & (0.1, 2) & (0.1, 2) & (0, 2) & (0, 2) & (2, 4) \\
$l_\text{tr}$ (cm) & $r=$(10, 65) & $r=$(10, 65) & $r=$(10, 65) & $r=$(10, 65) & $z=$(45, 70)\\
\bottomrule
\end{tabular}
\caption{Selection criteria for the conventional and track-based (TB)  analyses presented in this
  paper, with $p_{T,\rm tr}$ and $p_{\rm tr}$ referring to the
  transverse and total momentum, respectively, of the  charged track with length
  $l_{\rm tr}$. Unless otherwise noted all values given are minimum
  values for the quantity in question.}
\label{tab:cuts}
\end{table}

The sensitivity in this search is controlled by three parameters, the mass of the decaying charged particle
$m_{\chi}$, its nominal decay length,$c\tau$, and the cross section for
production of the charged state.  The event kinematics and analysis
efficiencies are fixed by the first two, we will treat these as free
parameters and scan over them. The cross section sets the
overall normalisation, and hence the number of events and search
significances, and is a fixed function of mass for a specific choice of quantum numbers and interactions
for the dark sector particles.  We will present our results at two different
levels of specificity: at the most general level we remain agnostic
about the origin and properties of the dark sector, and provide
information on the production cross section necessary to produce a
fixed number of events (corresponding to an estimated $5\sigma$
significance) for each mass and decay length, for ease
of recasting.  We will then use the simplified model described in Section~\ref{sec:SimpMod} to set
the production rate, and translate these results into projected sensitivities
in the $(m_\chi,c\tau)$ plane.  Throughout this work, simulations will be carried out
using this simplified model, although as stated above, we expect the
analysis efficiencies to be rather insensitive to the details thereof.

\subsection{Conventional analysis}
\label{sec:Central}
\begin{figure}
\subfloat[]
{
\includegraphics[width=0.49\textwidth]{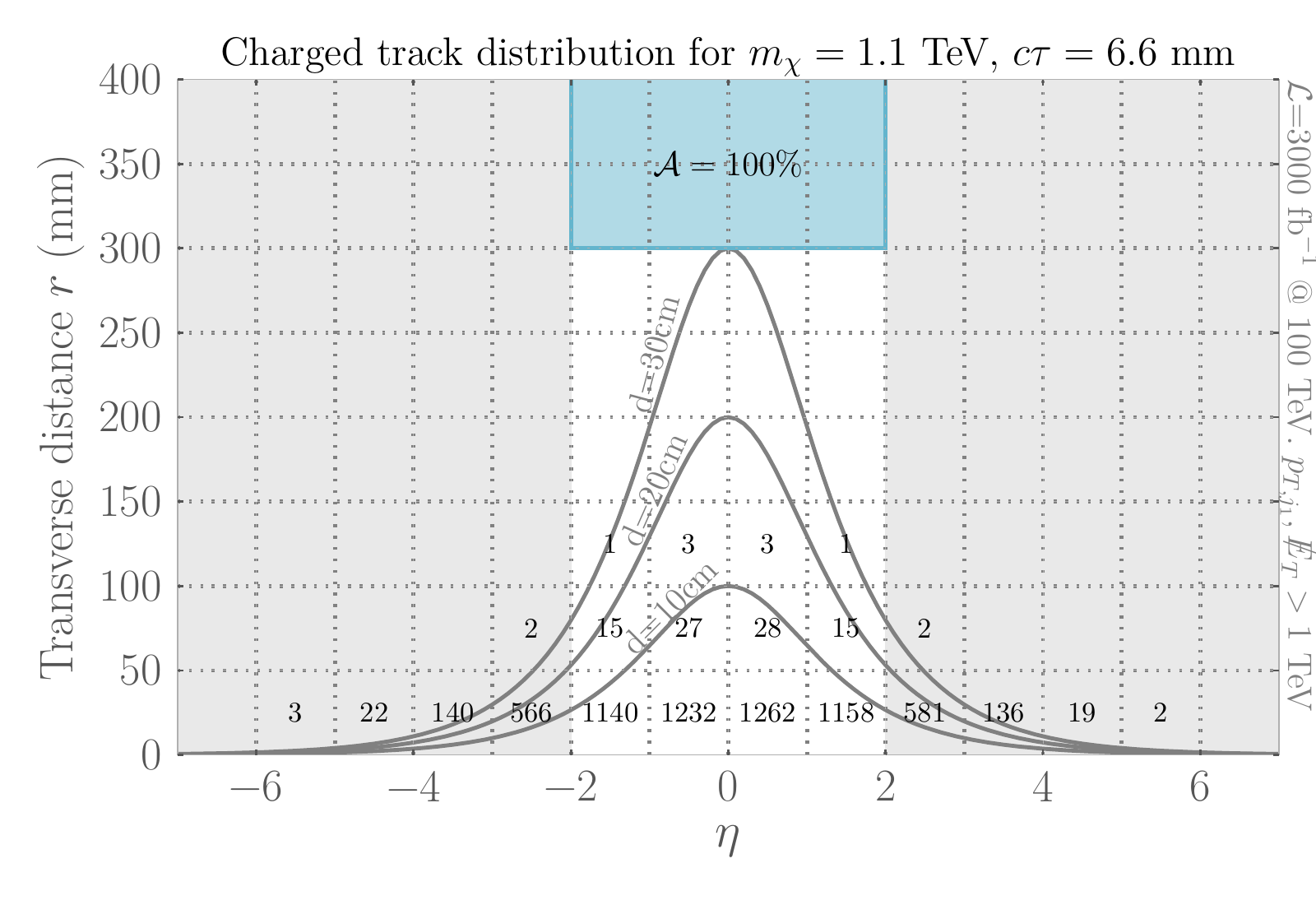}}
~
\subfloat[]
{
\includegraphics[width=0.49\textwidth]{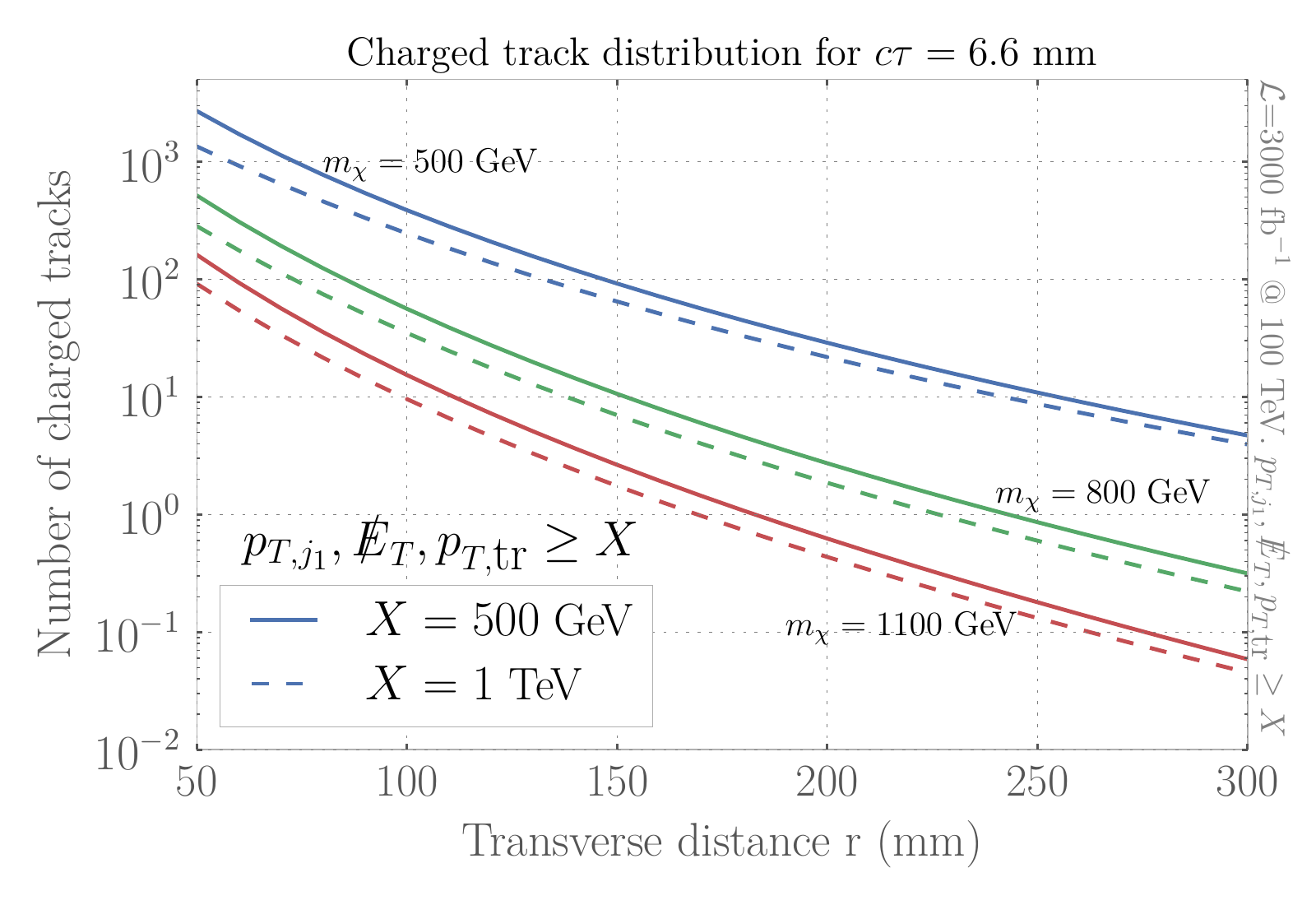}}
\caption{Distribution of number of charged tracks for weak-doublet
  fermion $\chi$ pair-produced at FCC-hh, as a function of $r$ (mm), (a) binned in $\eta$ and rounded down to the nearest integer, for $m_\chi=1.1$~TeV and $p_{T}(j_1),\slashed{E}_T, p_{T,\text{tr}}>1$~TeV and (b) integrated over $|\eta|\le 2$, for various masses and $p_{T}(j_1),\slashed{E}_T>X$. The lifetime is chosen equal to that of the corresponding pure Higgsino state in the MSSM, $c\tau=6.6$ mm.  The existing ATLAS search only has sensitivity for tracks that reach $r=30$ cm, shown as a cyan shaded region in the left panel (at CMS this number is slightly larger).}
\label{fig:EtaDistrib}
\end{figure}
We model this analysis on the disappearing charged track search
carried out by the ATLAS collaboration at 8~TeV~\cite{Aad:2013yna} with cuts roughly
scaled to 100~TeV centre-of-mass. We use the following pre-selection criteria:
\begin{itemize}
\item lepton veto;
\item leading jet $p_T>$ 1~TeV;
\item $\slashed{E}_T>$ 1~TeV;
\item $\Delta\phi_{\rm min}^{j,\slashed{E}_T}>$1.5 to reduce the contribution from hard jets faking MET.
\end{itemize}
In addition, ATLAS selects charged tracks that give at least 5 hits in the
inner tracker layers, of which 
3 are in the pixel tracker and 2 in the (double-sided) SCT, to facilitate a
good track reconstruction.  In order to satisfy this criterion a track must reach a
transverse distance $r$ of 30 cm, severely compromising the sensitivity to intermediate-lifetime
particles.  Instead, we will allow the radius at which 100\% efficiency is
reached for a disappearing charged track to vary, and show how this
affects the sensitivity for different masses and lifetimes of the
charged state.  The pseudorapidity range of the ATLAS analysis is
limited by the extent of the Transition Radiation Tracker (TRT), which
is used to veto long-lived tracks, but as we shall show later
increasing the $\eta$ range has little effect on the sensitivity of this analysis.  We will also
comment on the variation of the sensitivity with changes in
the leading jet $p_T$ and MET cuts.

Our charged track selection is as follows:
\begin{itemize}
\item $p_{T,{\rm track}}>1$~TeV;
\item transverse track length $r_\textrm{min}\le r_\textrm{track}\le 65$
  cm, with $r_\textrm{min}$ the minimum length necessary for a
  reliable track reconstruction;\footnote{We also assume the presence
    of a tracker at 65 cm that fulfils the role of the TRT.}
\item 0 $<|\eta_{\rm track}|<$ 2.
\end{itemize}

We show in figure~\ref{fig:EtaDistrib} the charged track distribution
with transverse distance from the beamline $r$.  In the left
panel (a), we bin the distribution in $\eta$ for
$m_\chi=1.1$~TeV and $p_T(j_1),\slashed{E}_T\ge 1$~TeV, while in the
right panel (b) we integrate the total number of charged tracks over
$|\eta|\le2$ , for three different masses, and
two different $p_T,\slashed{E}_T,p_{T,\text{tr}}$ cuts.  We take a nominal
decay length of 6.6 mm, which corresponds to the proper lifetime of
the charged component of 1.1~TeV pure Higgsino state, for
illustration.  The sensitivity to charged tracks in the current ATLAS
search begins at $r=30$ cm, shaded in cyan in the left panel.  It is
evident from the plots that TeV-scale charged states with sub-cm
$c\tau$ cannot make it out to large enough transverse distances in
order to be covered by existing searches. Were it instead possible
to achieve 100\% track reconstruction efficiency  within 10 cm of the beamline, one could have $\mathcal{O}$(10) events for weak-doublet masses comfortably up to 1~TeV.  

\begin{figure}
\centering
\subfloat[]
{%
\includegraphics[width=0.49\textwidth]{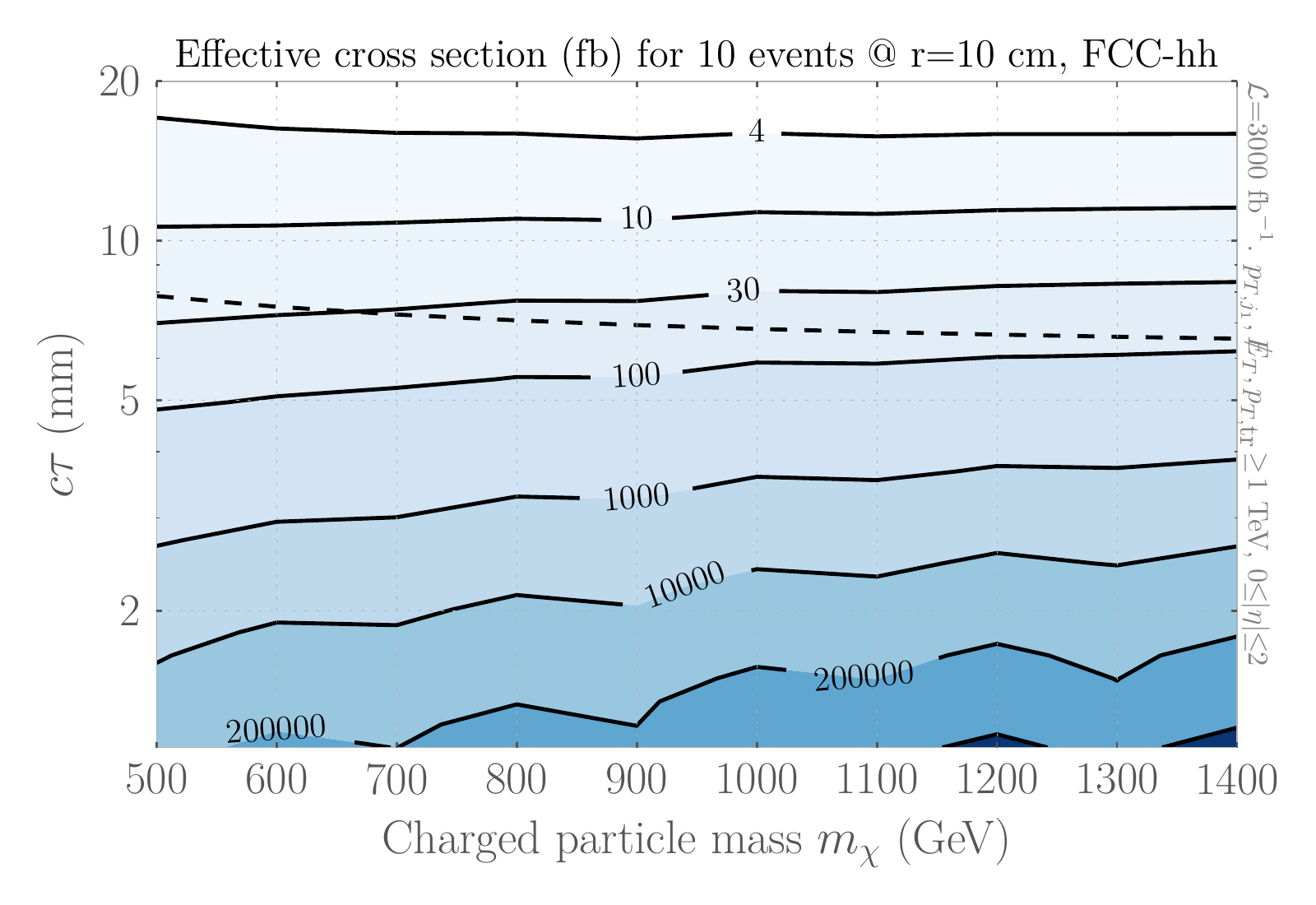}}
\subfloat[]
{
\includegraphics[width=0.49\textwidth]{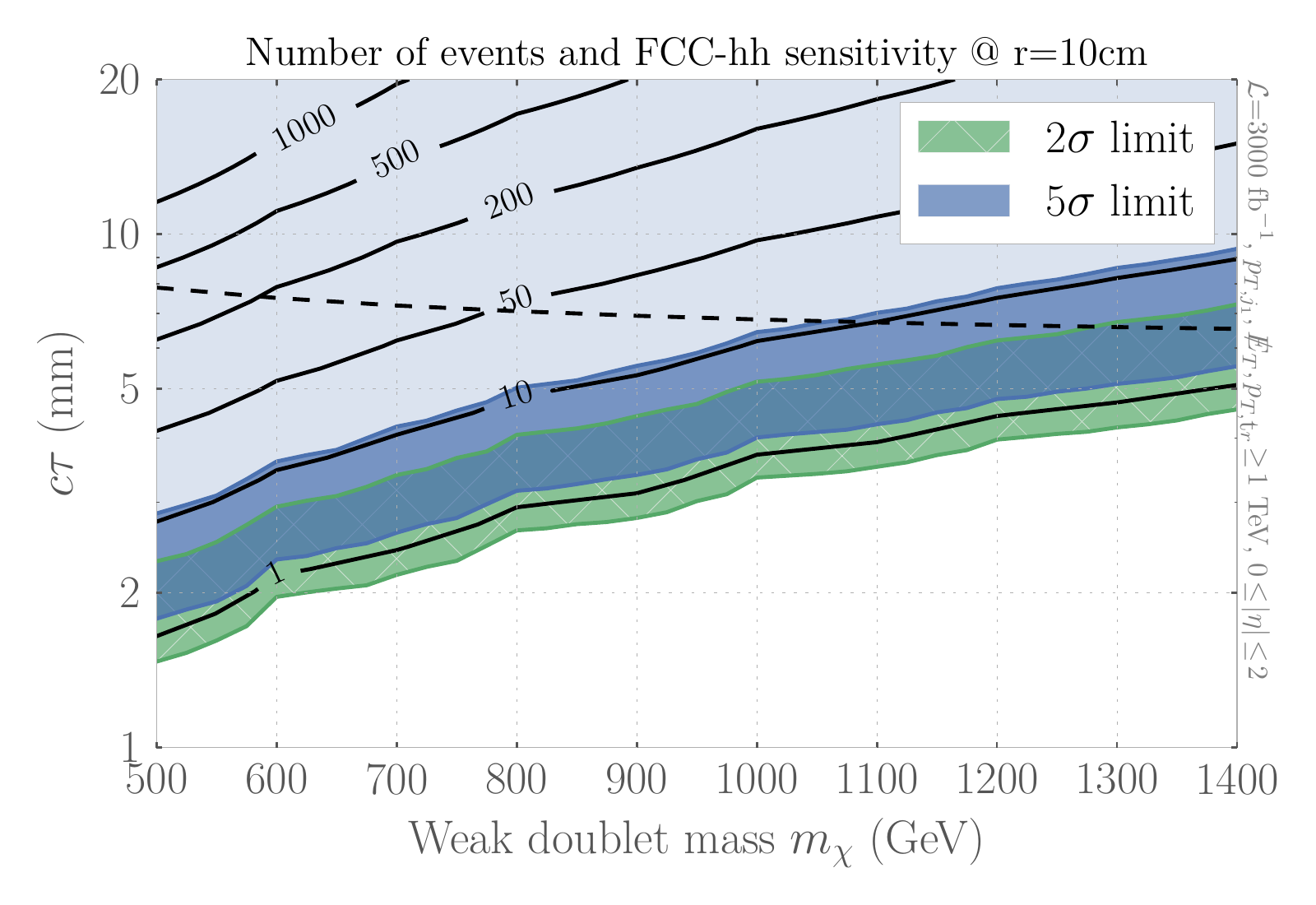}}
\caption{Results of conventional analysis: (a) Effective charged particle production cross section
  (definition in text)
  required in order to obtain 10 disappearing charged track events in
  conventional analysis at
  $r=10$ cm, and (b) number of disappearing charged tracks and sensitivity,
  normalized to the NLO pair-production cross section of a
  weak-doublet fermion with Dirac mass $m_\chi$ and
  nominal decay length $c\tau$.  The plots are for a $pp$ collider at
  $\sqrt{s}=100$~TeV with 3000 fb$^{-1}$ of integrated luminosity.
  The $c\tau$ corresponding to a pure Higgsino state is shown as a
  dotted line.  Superimposed onto the right panel (grey shaded region)
  is the FCC-hh sensitivity in this channel for a 50\%
  background systematic, with the estimated uncertainties in the $5\sigma$ ($2\sigma$) contours shaded in blue (green).}
\label{fig:XSNumber}
\end{figure}

Improving the track resolution at small $r$ does not seem outside the
realm of possibility, and one could envision various alternative
tracker arrangements that might achieve this.  Putting additional pixel barrel
layers in-between existing ones might suffice; as might angling the pixel
elements with respect to the beamline, as in the ATLAS Alpine upgrade
proposal, which could yield multiple hits from a single layer.
Since it is difficult to accurately assess the cost and
relative feasibility of such ideas, we make no further comment on the
matter, and simply assume in what follows that a reliable charged track resolution can
be achieved within a radial distance of 10 cm.  We then show in the
left panel of figure~\ref{fig:XSNumber} contours of the effective cross
section required to obtain 10 charged track events at $r=10$ cm, as a
function of $\mchi$ and $c \tau$.  The `effective' nature of the cross
section stems from the fact that pair-production of a weak
multiplet includes sub-processes with different numbers of charged
states, each of which has a different cross section, and is defined by
$\sigma_\textrm{eff} \mathcal{L}=N_{\chi^\pm}$, the number of
charged states produced.  This plot can then be used to
estimate the sensitivity for any scenario with direct pair-production
of a short-lived charged particle.\footnote{Caution must be exercised 
  in applying the results to charged particles produced in cascade
  decays, whose boost spectrum is instead set by the mass
  differences between the parent and daughter particles, and could be
  unlike that of a particle pair-produced at threshold.}  We
see for instance, that obtaining 10 charged tracks at
$r=10$ cm for TeV-scale charged particles and $c\tau$ between 5 and 10
mm would require a cross section of $\sim 100$ fb, in the correct
ball-park for weak production.

We also show in figure~\ref{fig:XSNumber}(b) contours of the total
number of charged tracks as function of $c\tau$, assuming a
100\%-efficient disappearing track selection at $r=10$ cm.  This is normalized to the NLO production cross section for a weak-doublet fermion with Dirac mass $m_\chi$, at a 100 TeV $pp$ collider with 3000 fb$^{-1}$ of integrated luminosity.  In both cases the $c\tau$ for a pure Higgsino state is shown as a dotted line.

Converting a number of tracks to a discovery/exclusion significance requires some knowledge of the size of SM backgrounds to this process.  There are no real backgrounds satisfying the analysis criteria.  Fake
backgrounds consist of interacting hadron tracks, leptons failing
identification criteria at low track $p_{T}$, and tracks with
mismeasured $p_T$ due to ``a high density of silicon hits, hadronic
interactions and scattering''\cite{Aad:2013yna} at large track
$p_{T}$.  These fakes are not well-described by Monte Carlo
simulations at the LHC at 8~TeV centre-of-mass. Instead, their $p_T$
spectra are fit to data in a `control' region and subtracted, rendering their extrapolation to 100~TeV rather difficult.  In addition their
composition and spectra are characteristic of the particular detector
in which they are measured (ATLAS in this instance), and a naive extrapolation
to a hypothetical detector for a 100~TeV hadron machine, with unknown properties, would be crude
at best.  Nevertheless we will make some attempt to do this.  First,
we assume that the fake backgrounds at FCC-hh have a similar composition and are
again dominated at high track $p_T$ by the mismeasured hadronic tracks
satisfying the ATLAS 8~TeV disappearing track selection. We assume the hadronic fakes satisfying our {\it
  modified} selection criteria retain the same scaling with track
$p_T$ as the original ($p_{T,{\rm
    track}}^{-a}$ with $a=1.78\pm 0.5$), with a floating overall normalization that parametrizes our uncertainty.  This normalization constant can be estimated using the scaling of some chosen process
with centre-of-mass energy.  Previous works~\cite{Low:2014cba,Cirelli:2014dsa} used
Standard Model $(Z\rightarrow\nu\nu)$ plus jets, the rate for this process scales with the product of
the quark and gluon PDFs.  In order to be maximally conservative,  we will also show the outcome using the scaling of SM multijets, with large fake MET.  This is glue-glue-initiated, and hence grows faster with energy.  We estimate the significance using the usual gaussian expression, 
\begin{equation}
{\rm Significance}=\frac{S}{\sqrt{B+\lambda^2 B^2}}
\end{equation}
for $S$ signal events and $B$ background events, and background
systematic uncertainty $\lambda=50\%$, using the difference between
the two methods of scaling the backgrounds as a measure of our total
background uncertainty.  We superimpose the resulting discovery
($5\sigma$) and exclusion ($2\sigma$) bands on to the event count
contours in figure~\ref{fig:XSNumber}(b).  The grey shaded region shows
the extent of the FCC sensitivity, with the width of the coloured
bands representing our estimated uncertainty in the significance
contours.  We see that the FCC with modified tracking can comfortably
cover $c\tau$ down to about 7 mm for $m_\chi\sim 1$~TeV.  The extent
of the sensitivity at larger masses depends on which estimate of the
backgrounds is closer to reality, with our optimistic projection extending discovery reach to $m_\chi=1.4$~TeV.

\subsection{Track-based (TB) analysis}
\label{sec:TB}
\begin{figure}
\centering
\subfloat[]
{%
\includegraphics[width=0.49\textwidth]{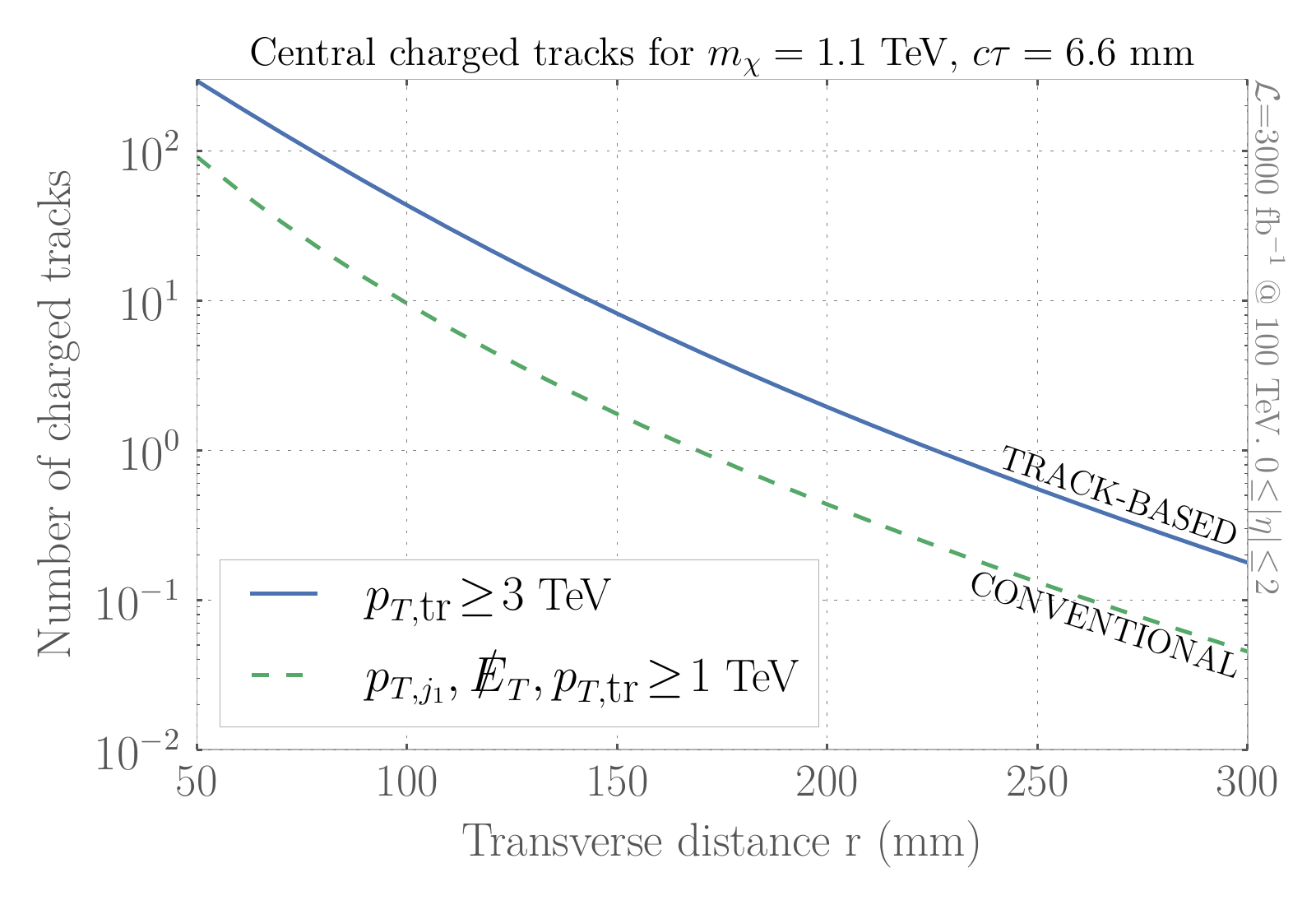}}
%~
\subfloat[]
{%
\includegraphics[width=0.49\textwidth]{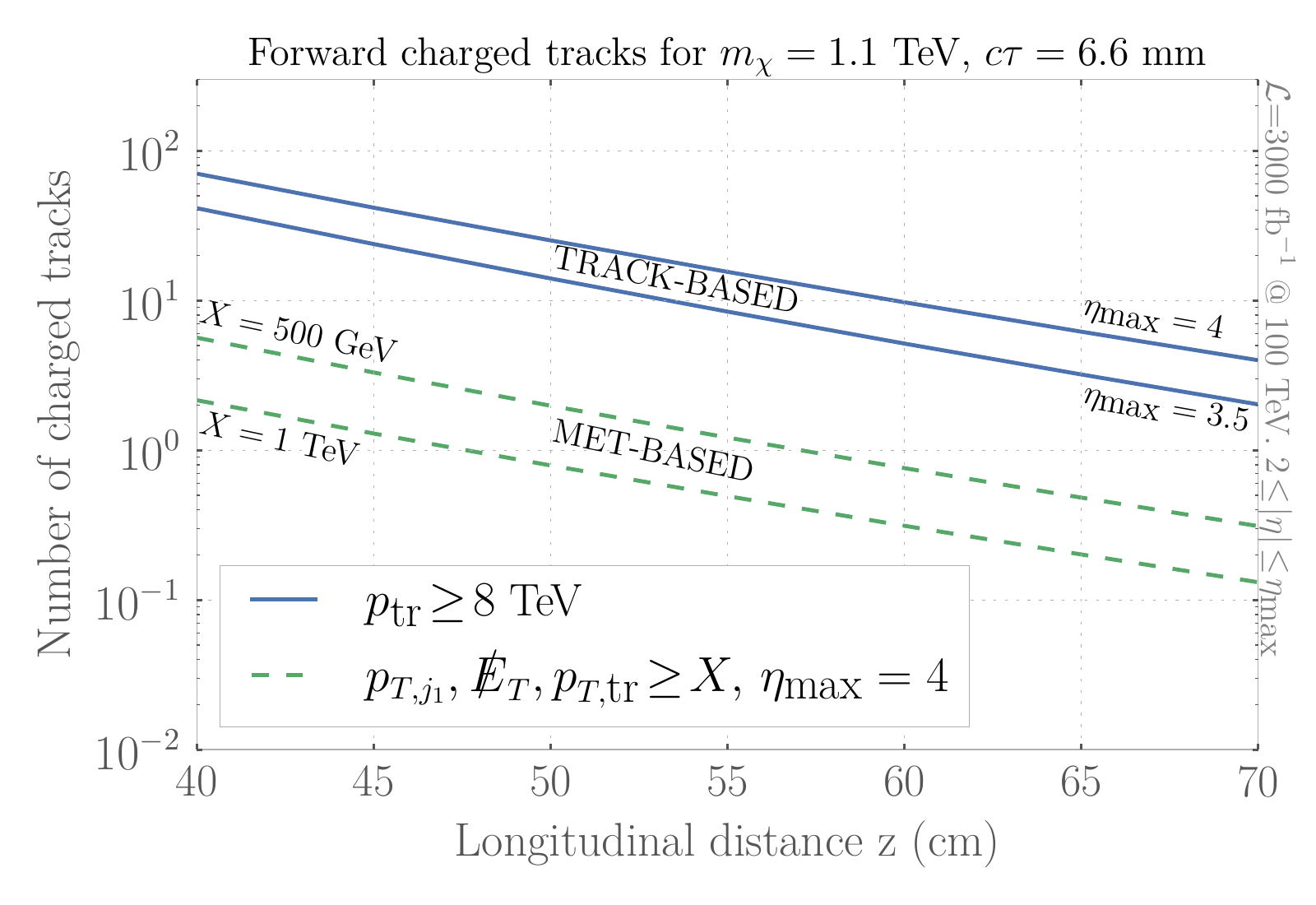}}
\caption{Number of charged tracks in track-based (TB) analysis for
  reference values $m_\chi=1.1$~TeV, $c\tau=6.6$~mm, in (a)
  central region as a function of transverse distance $r$ from the
  beamline, with the number of tracks in the conventional analysis
  shown for comparison, and (b) forward region as a function of the
  longitudinal distance $z$ from the interaction point, for two
  different values of the pseudorapidity extent.  We also show the
  number of charged tracks available in the forward direction using
  standard jet $p_T$ and MET cuts.}
\label{fig:CompCentralForward}
\end{figure}
At the beginning of this section we argued that cuts in recoil
$p_T$ and MET work to our disadvantage since they exclude events
containing particles with the largest boost, which travel furthest
from the interaction point.  In this re-designed analysis we will
dispense with recoil $p_T$ and related quantities, both for the purposes of
triggering and analysis cuts.  Instead we will cut directly on the measured
track momentum, which
is a proxy for the charged particle's boost, and hence the average lab-frame track length, for a
fixed $c\tau$.  We will assume that a hard enough (multi-TeV) trigger
requirement on this quantity, perhaps in association with a
$dE/dx$-based selection, will reduce data storage rates and 
backgrounds to manageable levels. However, since it is impossible for
us to make a reliable estimate of the size of fake backgrounds satisfying these modified disappearing track criteria, we will simply compare the sensitivity of our track-based analysis to the more conventional one at the level of the number of charged tracks.

We perform our track-based analysis in two non-overlapping signal regions. The central
region $0\le|\eta|\le 2$ will contain the bulk of of the charged
tracks. In addition, we define a forward region $2\le|\eta|\le \eta_\textrm{max}$ for some $\eta_\textrm{max}$ to be
specified later, which contains fewer but highly-boosted charged states, as illustrated in
figure~\ref{fig:MomDistrib}(c). The former will share part of the
disappearing charged-track selection criteria of the conventional
analysis described above, but none of the event-selection, which
relied on hard additional radiation.  In the central track selection we require:
\begin{itemize}
\item $p_{T,\textrm{tr}}\ge 3$ TeV;
\item $0\le|\eta|\le 2$;
\item $10$ cm $ \le r_\textrm{tr}\le 65$ cm.
\end{itemize}
The charged track distribution as a function of $r$, for $m_\chi=1.1$~TeV, $c\tau=6.6$ mm is shown in
figure~\ref{fig:CompCentralForward}(a), as is the analogous
distribution for the conventional analysis.  We see that dropping the hard radiation
requirement increases the number of charged track events by a factor
of 4 at a transverse distance of 10 cm, yielding more than 40 events
for our reference value.  

We further defining a track-based forward region as follows:
\begin{itemize}
\item $p_\textrm{tr}\ge 8$ TeV  ;
\item $2\le|\eta|\le 4$ \, .
\end{itemize}
Now we study the charged track distribution as a function of
longitudinal distance $z$ from the interaction point, for two
different values of $\eta_\textrm{max}$ (see
figure~\ref{fig:CompCentralForward}(b)), with the smaller value giving a
factor of 2 decrease in charged track yield.  By contrast, there are
an order of magnitude fewer tracks available if the traditional
$p_{T,\textrm{j}}$/MET-based cuts are maintained.

\begin{figure}
\begin{minipage}[c]{0.49\textwidth}
\center
\includegraphics[width=\textwidth]{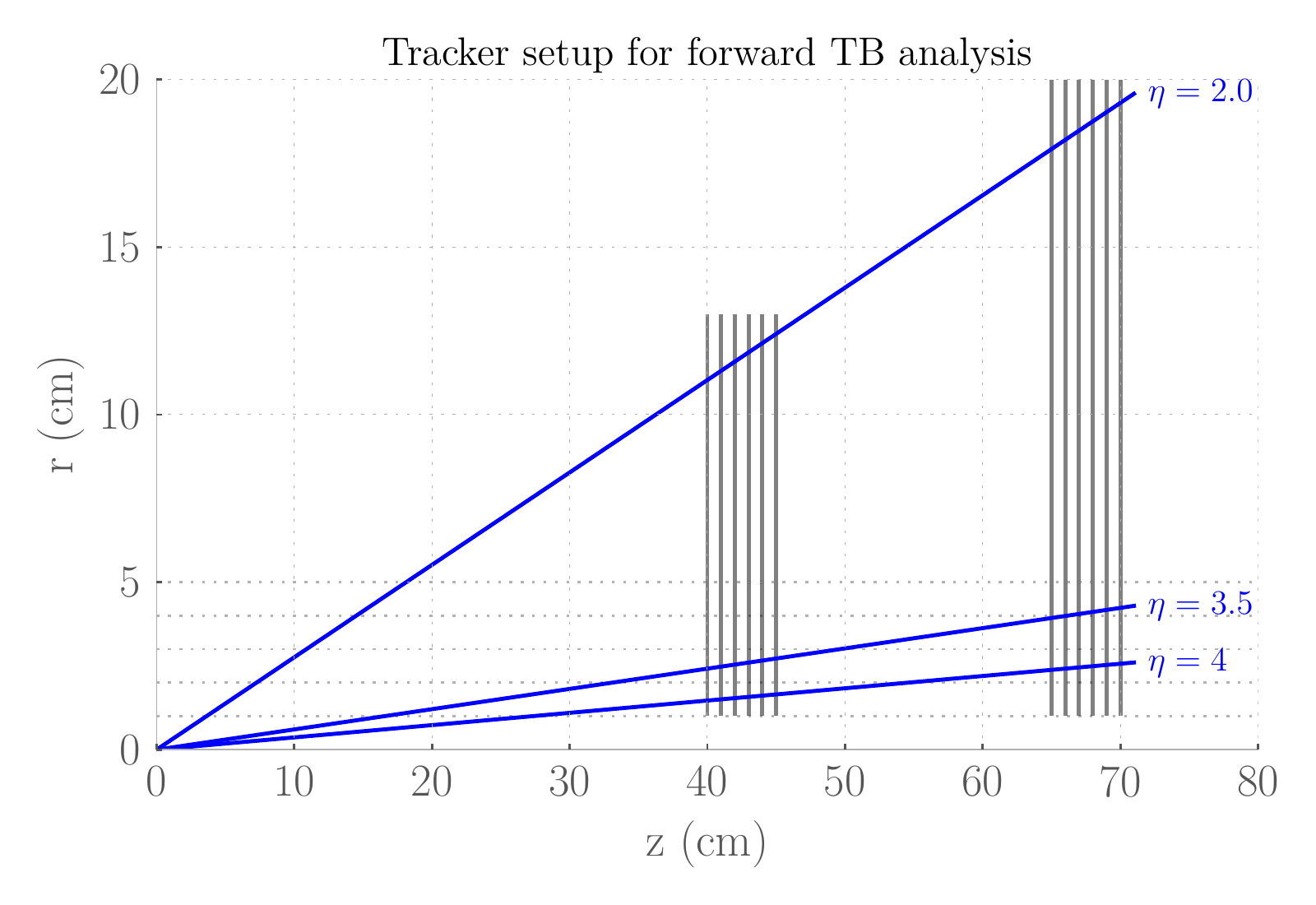}
\end{minipage}
~
\begin{minipage}[c]{0.46\textwidth}
\caption{Sketch of the proposed forward tracker setup, consisting of two
endcaps, each 5cm wide, located at $z=40$ and 65 cm, providing
hermetic coverage of the forward region $2\le|\eta|\le 4$, for
example. We assume any track reaching the far side of the near endcap
has yielded sufficient tracker hits to allow a reliable track
reconstruction, and, veto on tracks traversing the far endcap.}
\label{fig:FwdDetector}
\end{minipage}
\end{figure}
A complete definition of the forward signal region requires a concrete proposal for a
tracker that can successfully identify disappearing
charged tracks in the forward direction.  We have in mind a setup
similar to figure~\ref{fig:FwdDetector} with four tracking endcaps (two at each end) 
providing hermetic coverage of the forward region $2\le|\eta|\le 4$,
the far sides of which we take to be located at $z=45$ and $70$.
Although the endcaps, as illustrated, have a width of 5 cm, the precise value
is irrelevant provided we can assume any track reaching the far side of
the near endcap has left enough tracker hits to give a reliable track
reconstruction.  We also veto any charged track that traverses the far
endcap, which then plays the role of the TRT in the 8~TeV ATLAS
analysis, allowing us to exclude long-lived charged particles.~\footnote{Note 
that coverage to $\eta=4$ for the chosen endcap width mandates the
presence of tracker material within $r= 1.5$ cm of the beamline at
$z=40$ cm. This requirement can be relaxed by reducing the pseudorapidity
coverage or by moving the near endcap further from the interaction
point, at the cost of decreasing the sensitivity.}  We can then add to the TB forward track-selection criteria the
following requirement on track length: 45 cm $\le z_\textrm{tr}\le 70$
cm.

\begin{figure}
\centering
\subfloat[]{%
\includegraphics[width=0.5\textwidth]{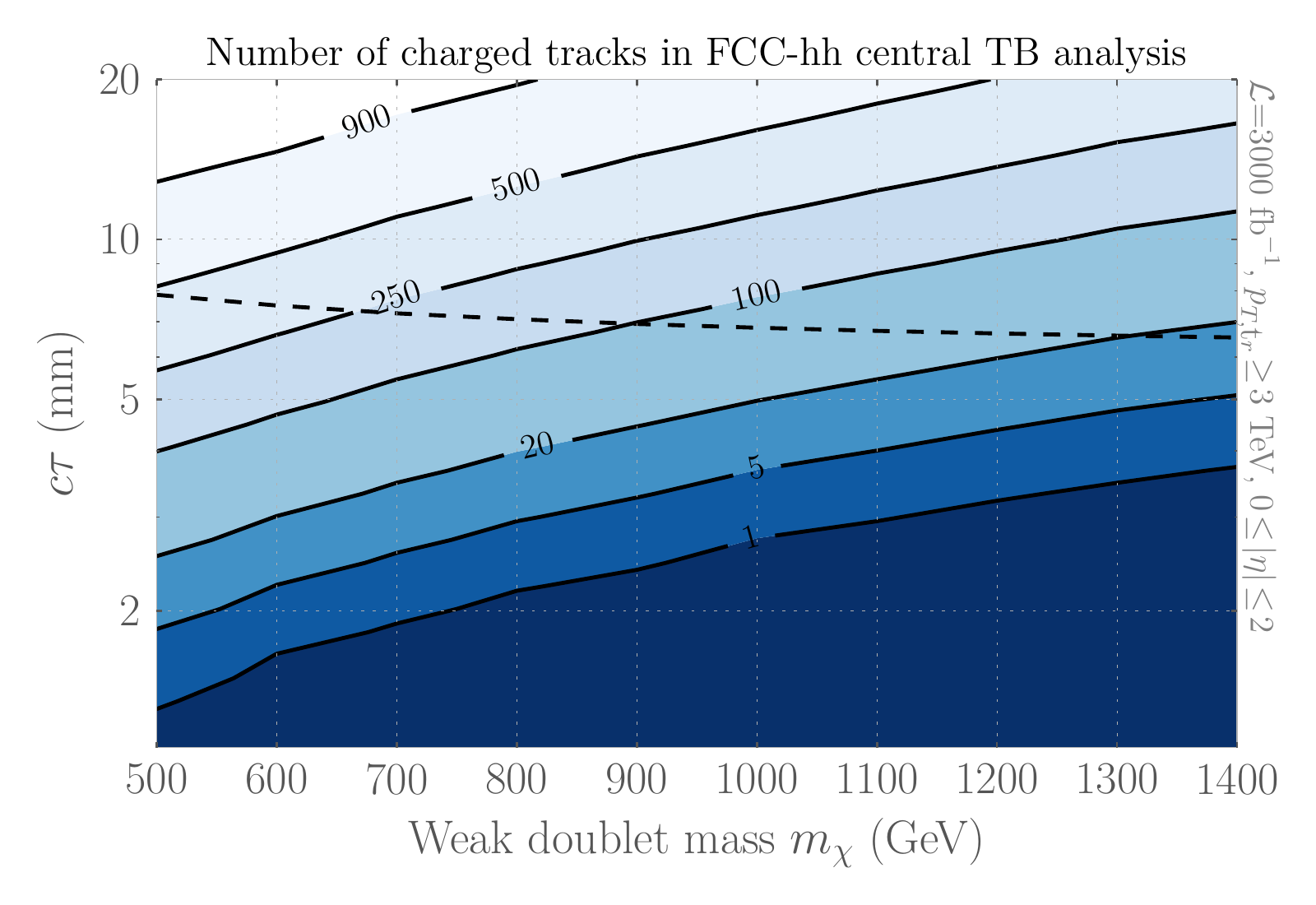}}
~
\subfloat[]{%
\includegraphics[width=0.5\textwidth]{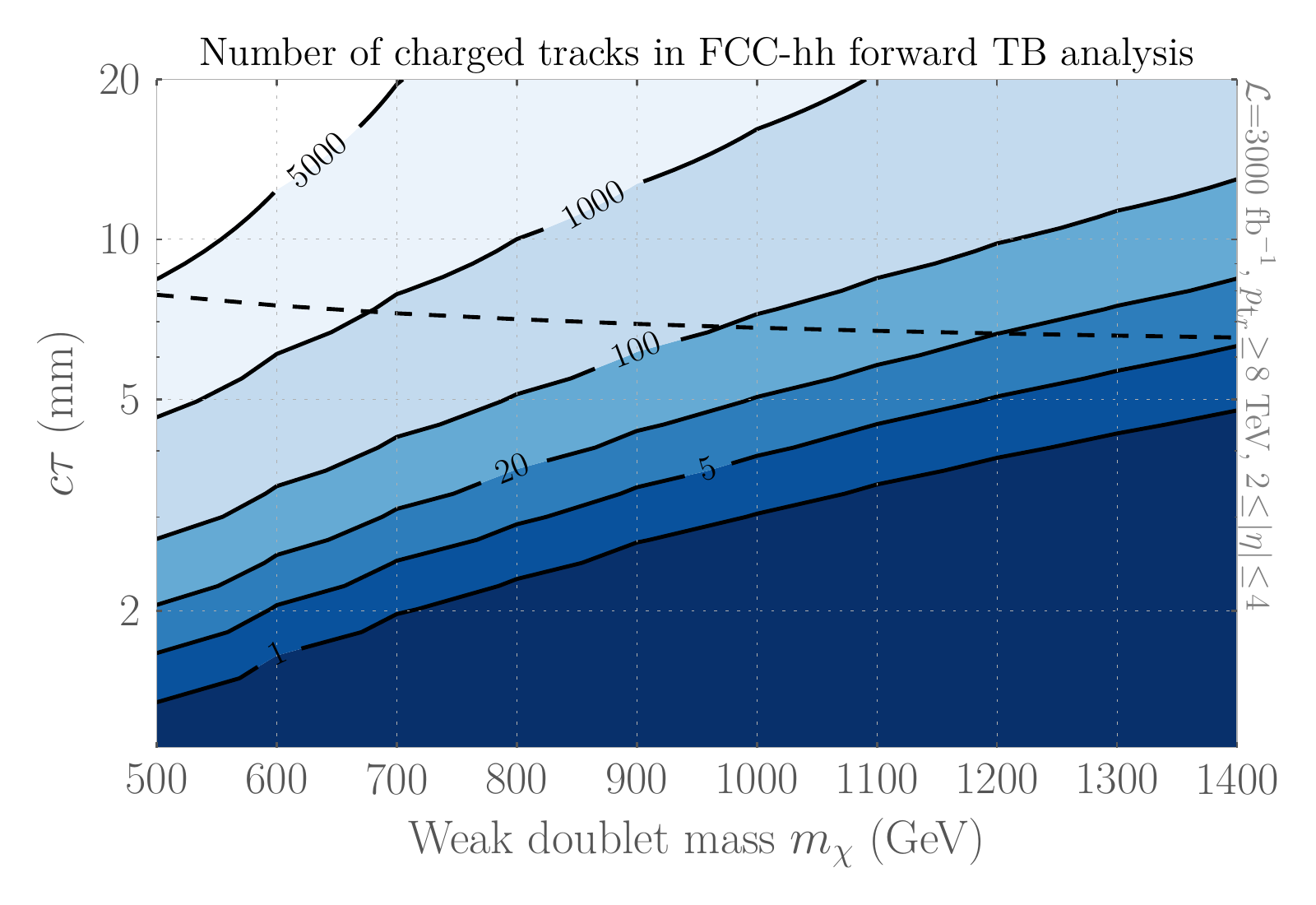}}
\caption{ Number of charged tracks in (a) central TB analysis 
(b) forward TB analysis,
 normalized to the NLO cross section of a pair-produced weak-doublet
 fermion with Dirac mass $m_\chi$ and nominal decay length $c\tau$. The $c\tau$ corresponding to a pure Higgsino state is shown as a dashed line.}
\label{fig:NPie}
\end{figure}

We display in figure~\ref{fig:NPie} (a) and (b) contours of the total number of
tracks in the TB central and forward
analyses respectively, as a function of
the weak-doublet mass $m_\chi$ and nominal decay length $c\tau$.  As
before the dashed line corresponds to the $c\tau$ for a pure Higgsino
state. Each of these track-based searches gives a
four-fold increase in the number of signal events relative to the
conventional analysis.

\subsection{LHC14-HL Sensitivity}

\begin{figure}
\centering
\subfloat[]{%
\includegraphics[width=0.5\textwidth]{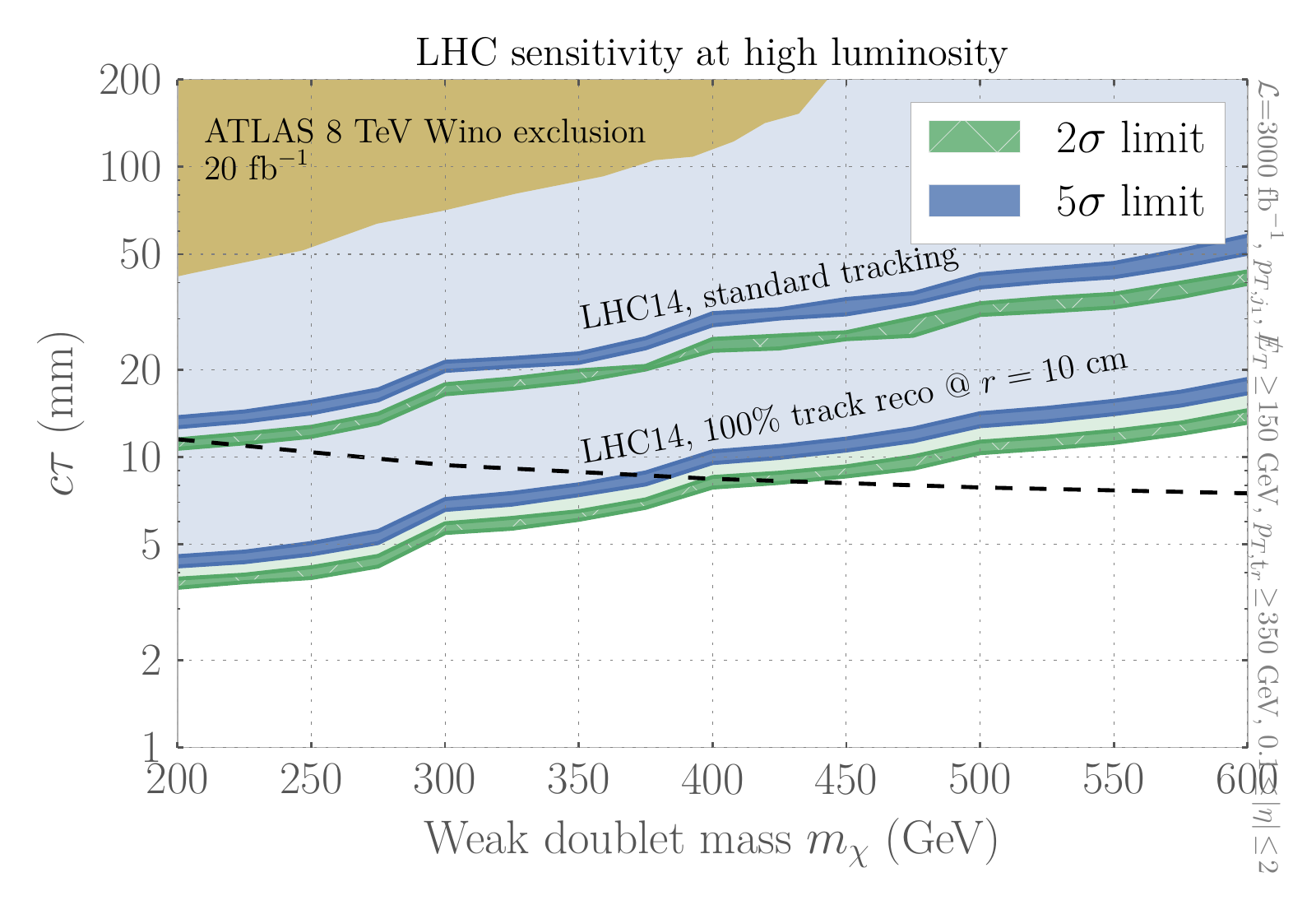}}
~
\subfloat[]{%
\includegraphics[width=0.5\textwidth]{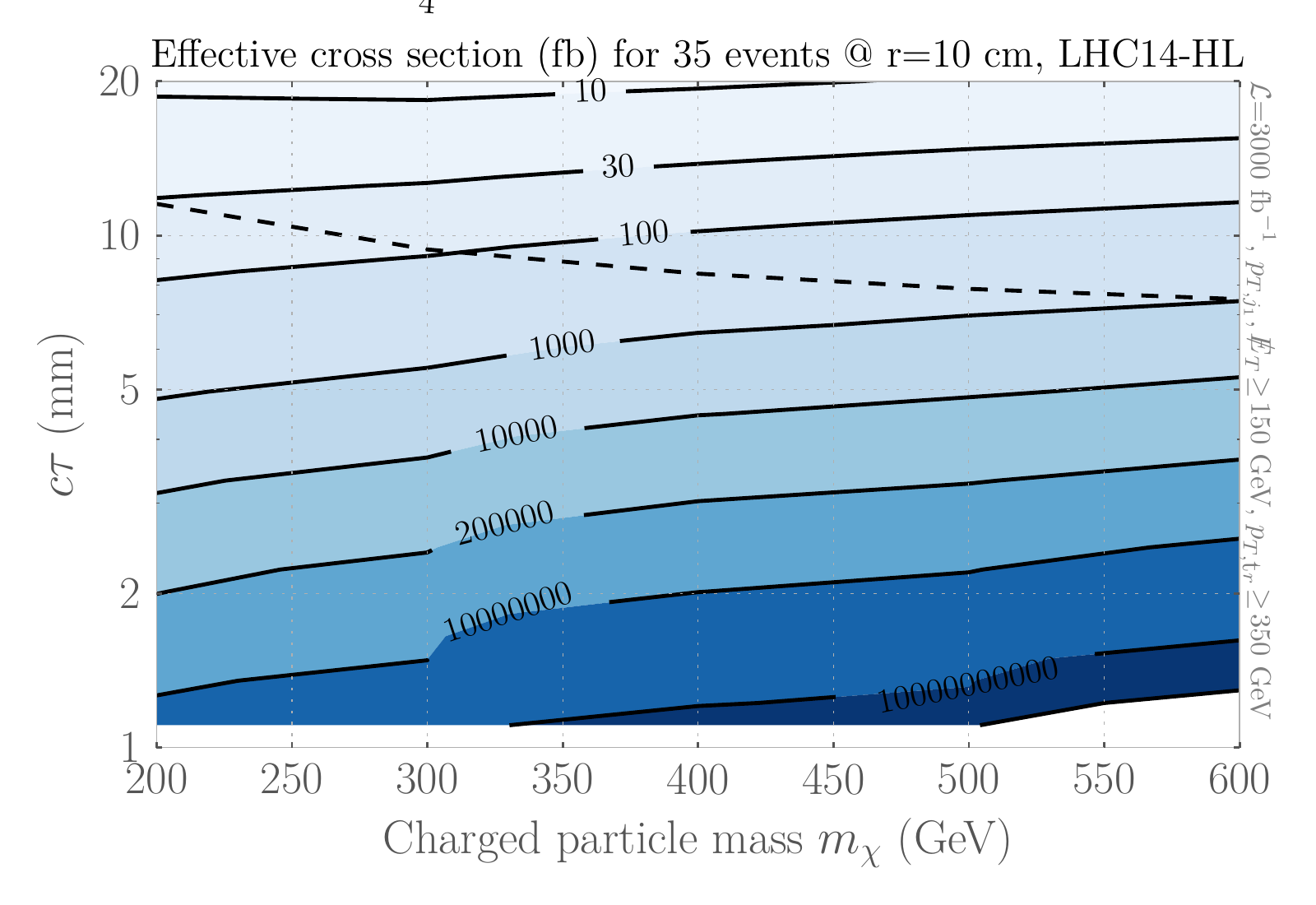}}\\
\subfloat[]{%
\includegraphics[width=0.5\textwidth]{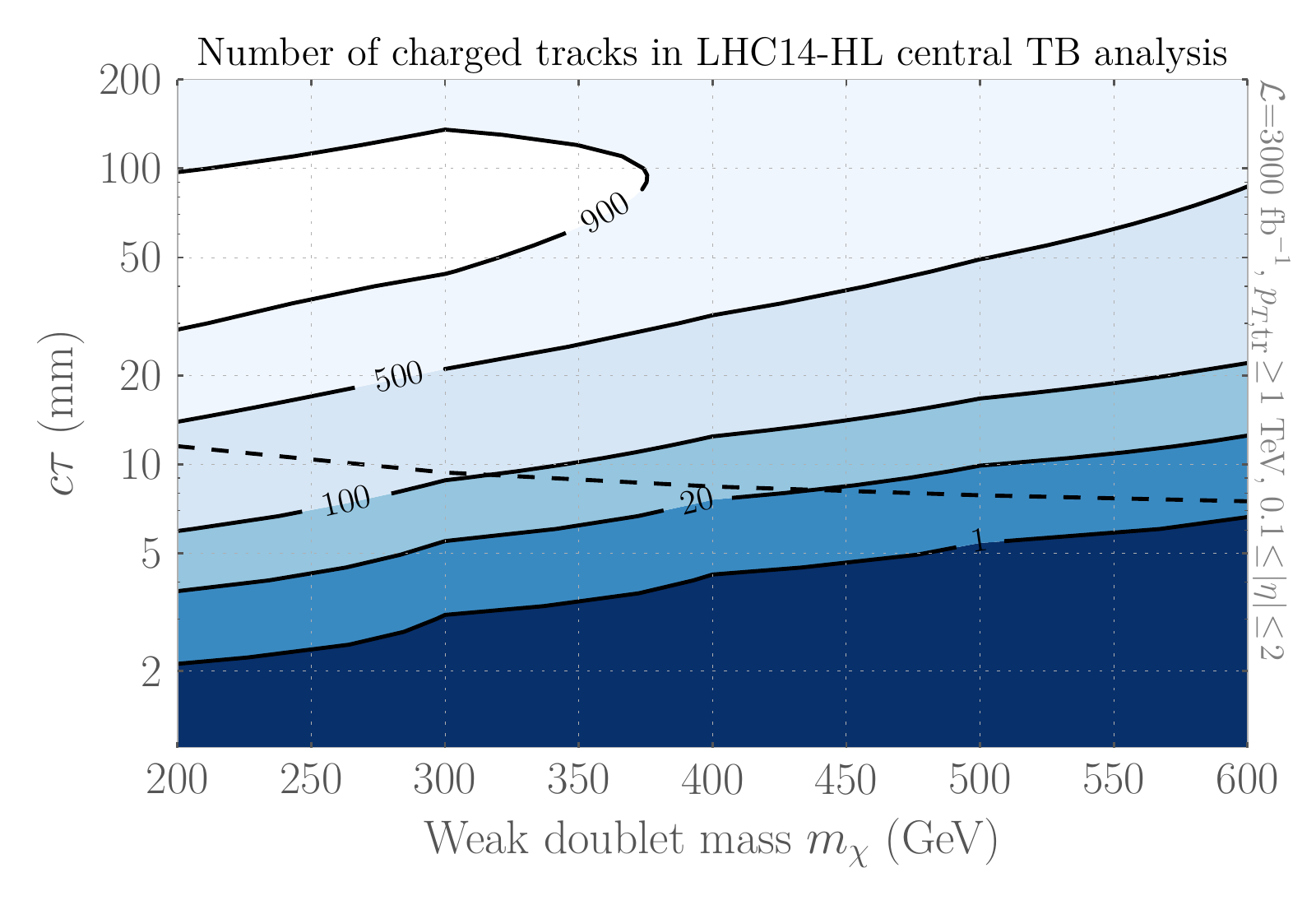}}
~
\subfloat[]{%
\includegraphics[width=0.5\textwidth]{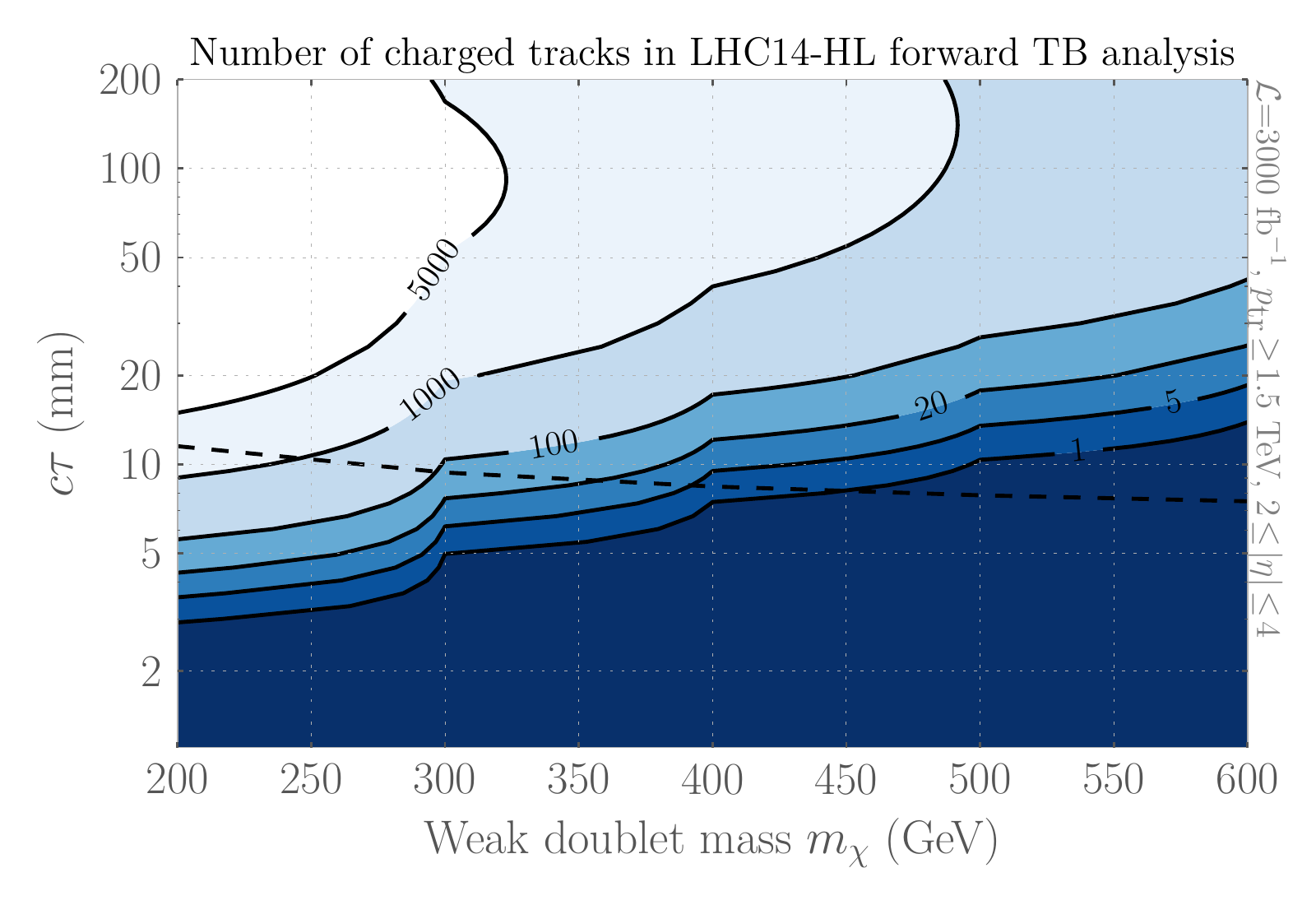}}\\
\caption{Reach for a high-luminosity run of LHC14, with
  $\mathcal{L}=3000$ fb$^{-1}$ integrated luminosity in (a)
  conventional analysis with and without (specified) improvements in tracker
  granularity close to the beamline.  We assume 50\% background systematics
, with estimated uncertainties (details in text) in the $5\sigma$ exclusion ($2\sigma$ discovery)
  contours shaded in blue (green). The yellow shaded region corresponds
  to the ATLAS 8~TeV limit for pure Wino states. (b) Effective cross
  section (defined in text) required for 35 charged-track events in
  conventional analysis at LHC14-HL, corresponding to a (conservative)
  $5\sigma$ discovery reach. (c) Number of disappearing charged tracks
  satisfying selection criteria in central TB analysis. (d) Number of
  disappearing charged tracks satisfying selection criteria in forward
  TB analysis. The dashed line corresponds to the nominal decay length for to pure Higgsino states.}
\label{fig:LHCReach}
\end{figure}

We can further compare these results to those obtainable at a
high-luminosity 14-TeV run of the LHC, with 3000 fb$^{-1}$ integrated
luminosity, assuming similar modifications are possible via detector upgrades. Our
selection criteria for the conventional analysis, now scaled to 14~TeV centre-of-mass, are:
\begin{itemize}
\item $p_{T,j_1},\,\slashed{E}_T\ge 150$~GeV;
\item $p_{T,{\textrm{tr}}}\ge 350$~GeV;
\end{itemize}
with the requirement on the charged track geometry unchanged from section~\ref{sec:Central}
above. For the central and forward track-based analyses we simply modify
the track momentum cuts of the FCC-hh version, to $p_{T,\textrm{tr}}\ge 1$~TeV and
$p_\textrm{tr}\ge 1.5$~TeV, respectively.  Our results, normalized
to the NLO cross section at the LHC14 from figure~\ref{fig:NLOXS}(a),  are shown in
figure~\ref{fig:LHCReach}. The backgrounds for the conventional analysis,
panel (a), were estimated as before, with the ATLAS 8~TeV limits on a pure Wino
state in the same channel are shaded in yellow. Recall that that the production cross section of the Wino
is larger than that of the Higgsino by a weak group Casimir factor.   We also
show our LHC14-HL projection with unchanged tracking capability in the same plot for
reference.  
We see that increasing the tracker granularity below $r=10$ cm as proposed above can result in a
factor of 3 increase in sensitivity at LHC14-HL,
extending the discovery reach down to $c\tau\sim 20$ cm for
$m_\chi=600$~GeV.
In this lower-energy environment the central TB analysis (figure~\ref{fig:LHCReach}(c)) still gives a two-fold enhancement to the number of
signal events. The forward TB analysis (Panel (d)) however is not as effective due
to the smaller overall boost, improving the charged-track yield by less than 30 \%.

A quantitative assessment of the negative impact on the TB signal
region of radiated-jet $p_T$ cuts can be found in Appendix~\ref{sec:ptcuts}.
Note that all our analyses above are simple cut-and-counts, un-optimized for the
range of mass scales they are applied to.  As a result our conclusions can be thought of as rather
conservative, with a dedicated optimization expected to yield improved
sensitivity over the entire mass range.

%-----------------------------------------------------------------------------
\section{Conclusions}
\label{sec:conclu}
%-----------------------------------------------------------------------------
One reason why we have seen no evidence of New Physics at the LHC could be that the lowest-lying new states have compressed spectra.  If these new states form part of a dark sector, with splittings so small that the SM by-products of decays within the sector become too soft to be detected, we have few handles to use in searching for such scenarios at hadron colliders. If the compressed dark sector contains a charged component, the disappearing track of this charged particle can be used in searches.  Such disappearing tracks will satisfy modified tracking criteria, and building in sensitivity to this type of exotic signature at future detectors will maximize our chances of closing the net for New Physics at future colliders.   

In this work we focused on extending the sensitivity of a future 100~TeV proton-proton collider to disappearing
charged tracks with sub-cm $c\tau$, by modifying detector specifications. While our motivation partly arises from covering TeV-scale masses (inspired by pure-Higgsino dark matter) our results are very general and apply to a variety of dark sector models.

We first explored the effect of small modifications to the tracker
setup on the outcome of
conventional searches.  While current searches are insensitive to
tracks that fall shy of 30 cm in the transverse direction, a reliable
reconstruction of charged tracks within 10 cm of the beampipe
would increase the reach in $c\tau$ by a factor of 3 at the
high-luminosity run of LHC14, and a further factor of 5 at FCC-hh,
effectively testing nominal decay lengths of a few millimetres.

Furthermore our simulations showed that employing hard cuts on the transverse momentum of SM radiation (and the related
quantity $\slashed{E}_T$) removes a significant fraction of the signal
events with the longest lab-frame decay lengths. In order to seize
these highly-boosted particles, we proposed selecting events by
requesting large overall momentum of a charged track, thus removing
the need for additional radiation to discriminate the event. We 
explored two different setups, relying on the use of 
a) the central region,  $0 < | \eta | < 2$,  with
modifications in tracking resolution as per our
`conventional' analysis and b) the
forward region, $2 < | \eta | < 4$ with additional tracking endcaps and a charged particle boost factor $\gamma \beta$ about twice that of the central region. As both regions
are expected to be subject to different backgrounds, we refrain from performing a
sensitivity estimation, but rather design our cuts so as to obtain
some modest number of signal events. 
Taking the conventional $p_T$ and $\slashed{E}_T$-based analysis as a
baseline, our new track-based search gives 5 times more observable
signal events in each of the central and forward regions, offering
tremendous potential for improvement.  A similar track-based search at
LHC-HL would increase the number of
events in the central region by a factor of 2, while the forward
region suffers from the lower available boost and is therefore less
promising.  While the potential increase in sensitivity is not as
dramatic as at the FCC-hh, studying the feasibility of implementing
track-based triggers at the LHC-HL still seems worthwhile. 

An alternative way of implementing the central TB analysis is as follows: imagine defining track-MET as the missing transverse energy computed using on tracks in the inner 10 cm of the tracker. By construction, signal events will have track-MET of the order of the chargino $p_T$, while at the calorimeter level the `calo-MET' is set by the $p_T$ of the leading jet. Background events by contrast, will not have a large discrepancy between track- and calo-MET, and hence the ratio of these two variables stands as a powerful discriminator between signal and background. While here this merely amounts to a rephrasing of the performed analysis, this ratio could also be employed to isolate more complex signals like Emerging Jets~\cite{Schwaller:2015gea}, where a large number of tracks may appear or disappear.

Here we have seen that track based searches with reduced dependence on inclusive kinematic variables are a promising direction for new physics searches, offering significant improvements in the experimental reach for larger masses and shorter lifetimes.  Combining the search strategies explored in this paper will further improve our sensitivity to charged states with picosecond lifetimes,allowing us to definitively discover, or exclude, a pure Higgsino thermal relic at 100~TeV.

\acknowledgments 
We would like to thank Benjamin Fuks, Roberto Franceschini, Frank
Krauss, Gavin Salam, Michele Selvaggi and Jure Zupan
for useful discussions, as well as the
Galileo Galilei Institute for Theoretical Physics, the Mainz Institute of Theoretical Physics (MITP) and the Theory group at CERN for their warm hospitality during the completion of this work.  RM was partially supported by ERC
grant 614577 ``HICCUP'' High Impact Cross Section Calculations for Ultimate
Precision, and the Swiss National Science Foundation under MHV grant 171330. P.S is supported by the DFG Cluster of Excellence PRISMA (EXC 1098). 

\begin{appendix}

\section{Decays and Lifetimes}
\label{sec:Decay}
\begin{figure}
\centering
\subfloat[]{%
\includegraphics[width=0.5\textwidth]{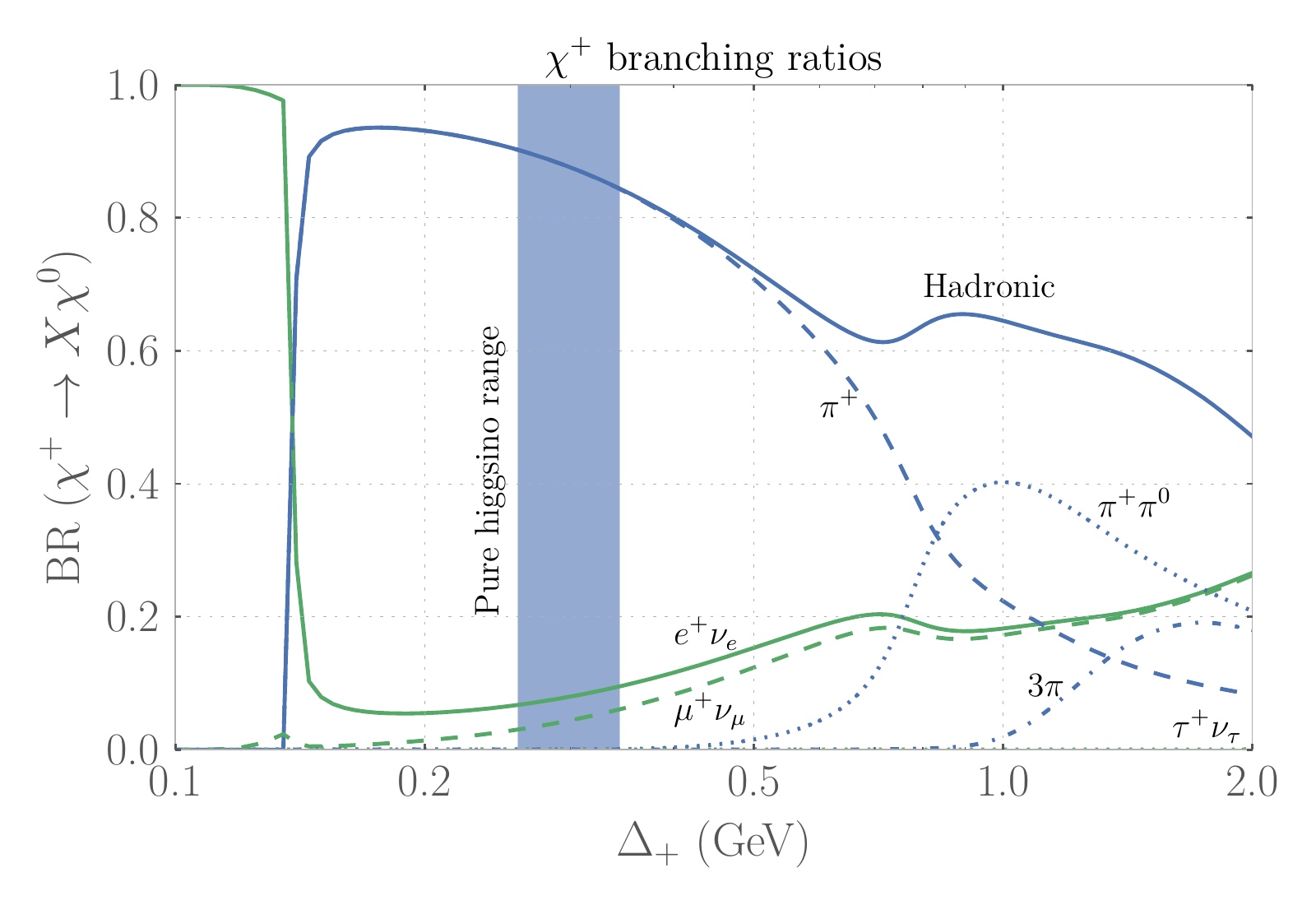}}
~
\subfloat[]{%
\includegraphics[width=0.5\textwidth]{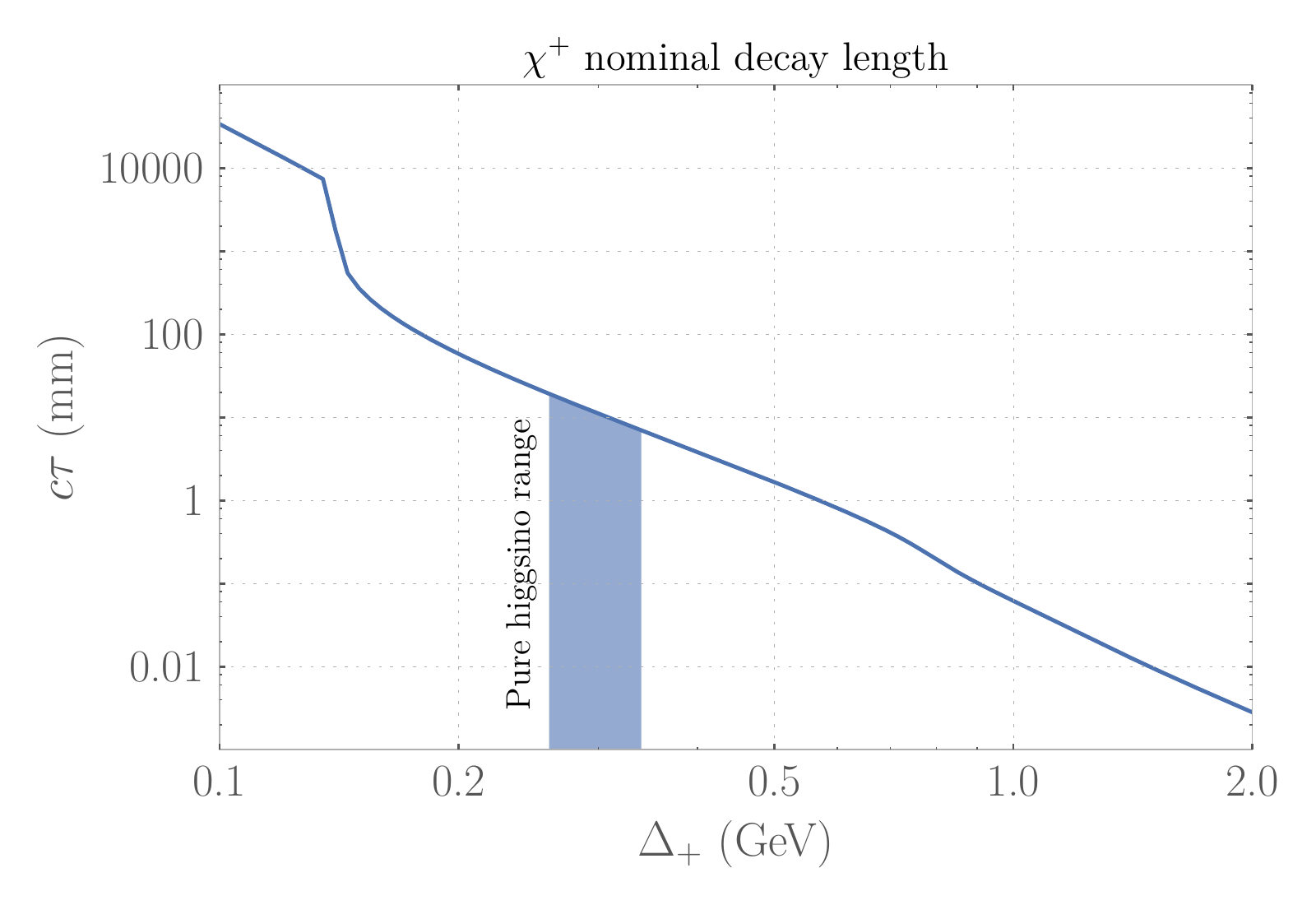}}\\
\caption{Branching ratios (left) and nominal decay length (right), for the charged component of an electroweak doublet with hypercharge \sfrac{1}{2}  and Dirac mass $\mchi = 1.1$~TeV, as a function of $\Dplus$, the splittingn between the charged and neutral states. Bote that the partial widths are highly insensitive to the overall mass scale.}
\label{fig:charBRs}
\end{figure}
For mass splittings that correspond to the lifetime range under
consideration, the charged state in the doublet will decay via
off-shell W bosons into the lighter neutral state, accompanied
predominantly by either a single soft pion or a lepton-neutrino
pair. The partial widths for the relevant channels were computed in
\cite{Thomas:1998wy}, and are given below for convenience.  They agree with the expressions found in Refs.~\cite{Chen:1996ap,Chen:1999yf} for the Higgsino in the pseudo-Dirac limit, where the Majorana splitting between the two neutralinos is taken much smaller than the chargino-neutralino splitting $\Delta^+$, and the decay widths into both neutralinos are summed.

The partial width into the neutral component plus two SM fermions, $f$ and $\bar{f}^{\prime}$, with $m_{f^{\prime}}\sim 0$ is:
\be
\label{eq:cha3F}
\Gamma (\chi^{-} \to \chi_{0} f \bar{f}^{\prime} ) = \frac{2 N_C G_F^2 (\Dplus)^5}{15 \pi^3} \sqrt{1-r_f^2}\mathcal{P}(r_f)
\ee
where $N_C$ is the number of fermion colours, $r_f=m_f/\Delta^+$ and 
\be
\mathcal{P}(r_f)=1-\frac{9}{2}r_f^2-4 r_f^4 +\frac{15 r_f^4}{2\sqrt{1-r_f^2}}\tanh^{-1}{\sqrt{1-r_f^2}}\;\;\;.
\ee
The partial width to a charged pion plus neutral component is given by:
\be
\label{eq:chapi}
\Gamma (\chi^{-} \to \chi^{0} \pi^{-} ) = \frac{2  G_F^2 f_{\pi}^2\cos^2{\theta_C}}{\pi} \Delta_+^3\left(1-\frac{r_\pi^2}{2}\right)\sqrt{1-r_\pi^2} 
\ee
where $f_{\pi} \approx 91.9$~GeV is the pion decay constant, $\theta_C$ is the Cabbibo angle and $r_\pi=m_\pi/\Delta_+$.

Expressions for subdominant modes including the decays into two and
three pions can be found in ref.~\cite{Chen:1996ap}. They are not
particularly illuminating but for completeness we have included them
in our numerical analysis. The overall impact of these channels is
small except for the region [0.7-2.5]~GeV where the two-pion channel
contributes ${\cal O} (10 \%)$ to the branching ratio.  We show in the
left panel of figure~\ref{fig:charBRs} the branching ratios of the
charged state $\chi^+$ into $\chi^0+X$ for $m_\chi=1.1$~TeV, while in
the right panel we show the corresponding nominal decay length $c
\tau$, with the blue shaded region corresponding to the viable range
of pure Higgsino parameter space.
Both the branching fractions and decay lengths are highly insensitive to the value of the Dirac mass in the limit $m_\chi\gg
\Delta_+$, but are instead very sensitive to the splitting, increasing
like $\sim\Delta_+^3$ in the entire region of interest. 
For values of the splitting given by the one-loop electroweak
contribution, the branching fraction to a single-pion final state is 97 \%. Due to the small decay length of the chargino we have restricted ourselves to $\Dplus < 1.5$~GeV. For this mass range we have explicitly checked that transitioning to a parton framework to compute this decays gives smaller values than the hadronic approach. Moving to heavier masses requires to perform a matching between both frameworks (see reference~\cite{Chen:1996ap}).

\section{Impact of transverse cuts on forward track-based analysis}
\begin{figure}[h!]
\centering
\subfloat[]{%
\includegraphics[width=0.5\textwidth]{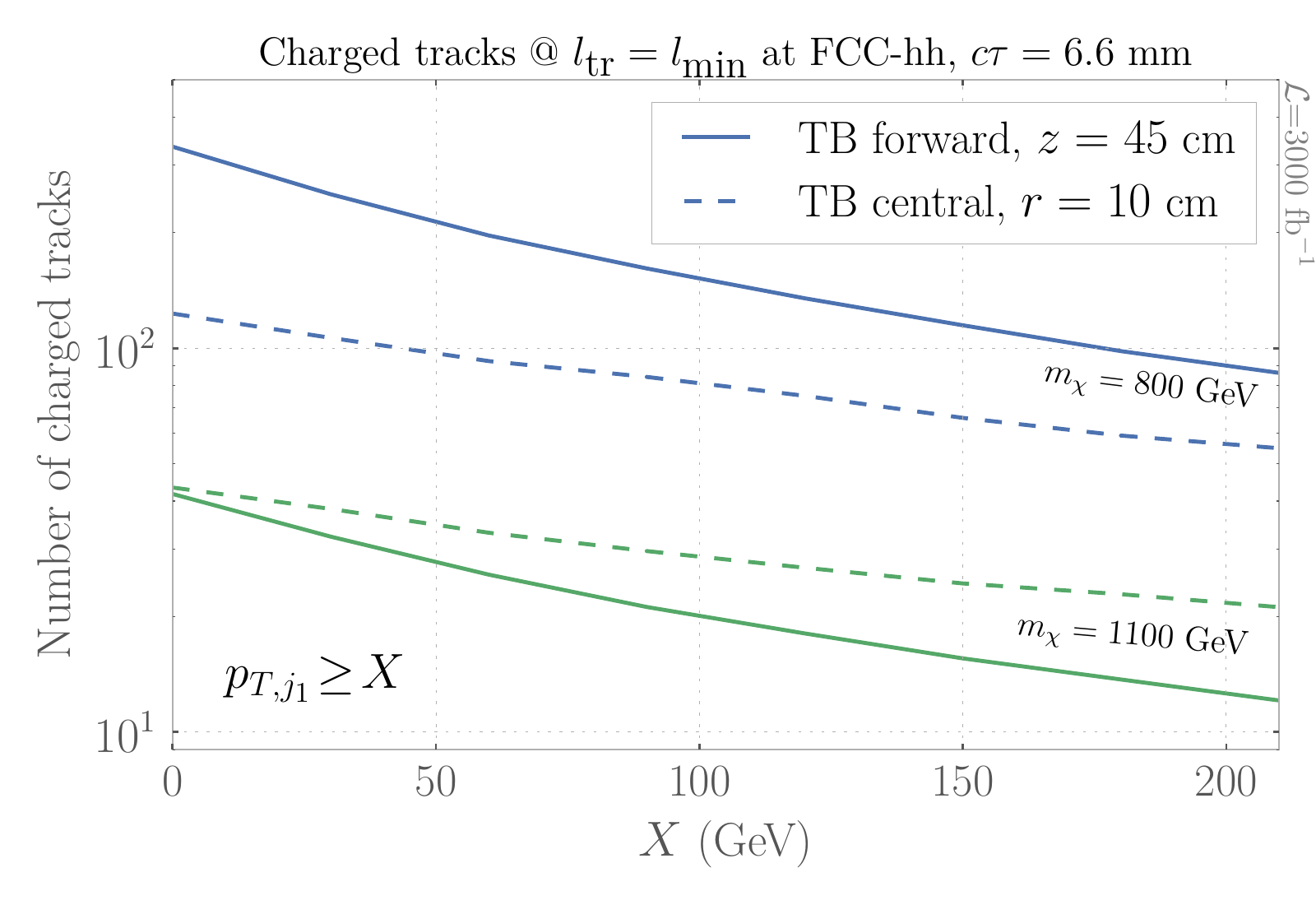}}
~
\subfloat[]{%
\includegraphics[width=0.5\textwidth]{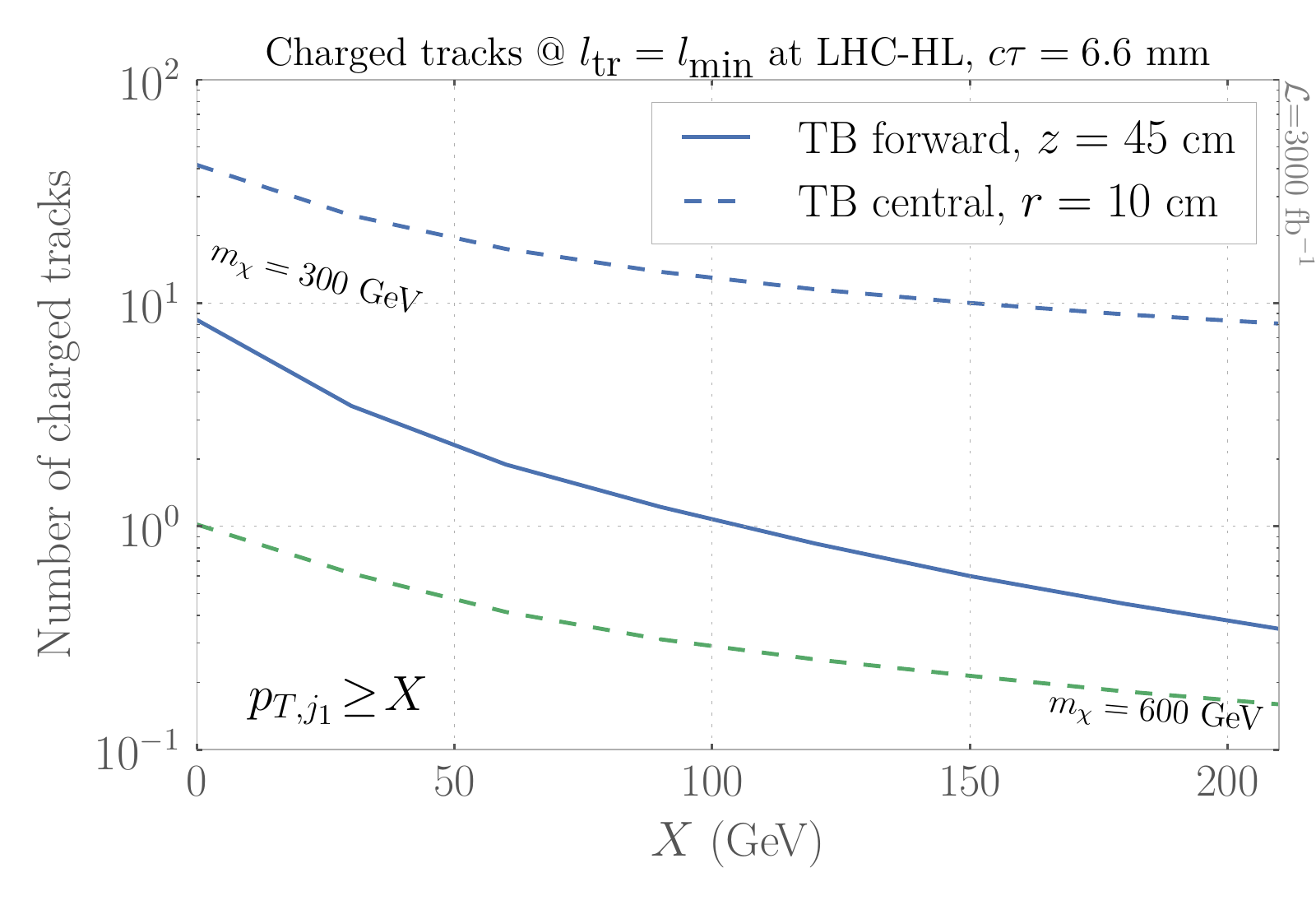}}
\caption{(a) Number of charged tracks in track-based analyses, at $z=45$ cm and $r=10$ cm in central region, as a function of a cut on the transverse momentum of the hardest recoil jet $p_{T,j_1}>X$. The total rate is normalized to the
  NLO production cross section for a weak doublet with mass
  1.1~TeV and $c\tau=6.6$~mm at (a) a 100 TeV $pp$ collider and (b) a 14 TeV $pp$ collider, with 3000 fb$^{-1}$ of integrated
  luminosity.}
\label{fig:ForwardTracks}
\end{figure}

While one of the main messages of this work was to propose the replacement of the transverse variables by track-related ones, it is still important to quantify the effect of performing transverse cuts since they could help keeping backgrounds under control.
 We thus quantify here the degradation of the signal in the TB searches due to transverse cuts by counting the number of charged tracks satisfying our selection criteria as a function of a soft cut on the $p_T$ of the leading jet, $p_{T,j1}\ge X$. The results for FCC-hh and LHC14-HL are presented in figure~\ref{fig:ForwardTracks}.
We see that even a very soft cut of $p_{T,j_1}\ge 100$~GeV at the FCC-hh decreases by
a factor of two the number of signal events available in the forward region, and keeps about 75 \% of the events in the central region. For the LHC14 we observe a similar behaviour for $p_{T,j_1}\ge 50$~GeV. As a result, one must  be cautious when using radiated jet $p_T$ in TB analyses, as even rather soft cuts sizably reduce the number of signal events.

\label{sec:ptcuts}
%

%------------------------------------------
\end{appendix}
\bibliographystyle{JHEP}
\bibliography{ChargedTrack}
\end{document}